 \documentclass[sn-apa]{sn-jnl}

\usepackage{graphicx}%
\usepackage{multirow}%
\usepackage{amsmath,amssymb,amsfonts}%
\usepackage{amsthm}%
\usepackage{mathrsfs}%
\usepackage[title]{appendix}%
\usepackage{xcolor}%
\usepackage{textcomp}%
\usepackage{manyfoot}%
\usepackage{booktabs}%
\usepackage{algorithm}%
\usepackage{algorithmicx}%
\usepackage{algpseudocode}%
\usepackage{listings}%
\usepackage{lineno}

\begin{document}

\title[Article Title]{Bayesian Multilevel Bivariate Spatial Modelling of Italian School Data} 


\author*[1]{\fnm{Leonardo} \sur{Cefalo} }\email{leonardo.cefalo@uniba.it} 

\author[1]{\fnm{Alessio} \sur{Pollice}}

\author[2]{\fnm{Virgilio} \sur{Gómez - Rubio} }

\affil*[1]{\orgdiv{Department of Economics and Finance}, \orgname{University of Bari Aldo Moro}, \orgaddress{\city{Bari}, \country{Italy}}}

\affil[2]{\orgdiv{Department of Mathematics, School of Industrial Engineering}, \orgname{University of Castilla - La Mancha}, \orgaddress{ \city{Albacete}, \country{Spain}}}


\abstract{This paper studies the relationship between the student's abilities in the second year of high school and the infrastructural endowment in all Italian municipalities, using spatial Bayesian modelling. Municipal student scores are obtained by averaging standardized and spatially homogeneous indicators of student outcomes provided by the Invalsi Institute for two subjects, Italian and Mathematics. Given the nature of the data, we employ a multilevel regression model assuming a bivariate Intrinsic Conditionally Autoregressive (ICAR) latent effect to explain the spatial variability and account for the correlation between the two subjects. Bayesian model estimation is obtained by the Integrated Nested Laplace Approximation (INLA), implemented in the \texttt{R-INLA} package. We find that alongside a significant association with the current state of school infrastructure and facilities, spatially structured latent effects are still necessary to explain the different student outcomes across municipalities.}

\keywords{Areal models, multivariate models, INLA, Invalsi}

\maketitle
\section{Introduction}

Territorial disparities in the Italian public education system are a severe and widely recognized issue. Exploratory analysis of school infrastructure endowment highlights a North-South divide encompassing several infrastructural dimensions, such as school accessibility and availability of learning and recreational spaces \citep{BDI}. The relevance of school infrastructure in learning processes is nowadays well documented \citep{WB}, hence it is reasonable to expect infrastructural disparities to be reflected in terms of student outcomes.

However, comparing student outcomes across a country's complex geography is not a trivial question. To this aim, a framework to define standardized and spatially homogeneous indicators has been developed by the OECD throughout the Programme for International Students Assessment \citep[PISA,][]{PISA}. In Italy, this task is attributed by law \citep{InvalsiLaw} to the Institute for the Evaluation of the Education System (Invalsi hereinafted). The Invalsi Institute runs an anonymous skill evaluation test whose attendance is mandatory across all public schools in Italy, regarding four subjects: Italian, Mathematics, English reading, and English listening. For each subject, the Invalsi elaborates a student-specific score through a methodology akin to the one employed by the PISA.

Indeed, territorial gaps in Invalsi scores are immediately evident and have been noticed to expand throughout the schooling process \citep{Invalsi2020}, the gap in high school scores being a matter of particular concern.  Considering analyses carried out at the individual (student) level for both PISA and Invalsi scores, a significant effect is often associated with North - Centre - South dummy variables  \citep[as in e.g. ][]{Giancola, Bratti, Agasisti, UniromaWP1} unless more explanatory variables regarding the labour market and socio-demographic dynamics are taken into account in relatively complex econometric models \citep{Bratti}. Additionally, \cite{Agasisti} partition the data set of Invalsi scores (last year of middle school) among Northern, Central and Southern Italy, running three different regression models, in which the intercepts range almost $11$ points apart in absence of explanatory variables and as far as almost $14$ points apart when some explanatory variables are introduced (at the time, Invalsi scores were expressed in a $[0-100]$ points range). A slightly different approach employs regression models with region-specific intercepts, allowing a higher amount of geographical information \citep{Matteucci} (in this case working with PISA scores); estimated intercepts display a clear territorial pattern, with all Northern regions exceeding the nationwide average and all Southern regions except for Apulia and Basilicata below it.

In this paper, we propose to model the average Invalsi scores for Italian municipalities using spatial analysis. We employ the Invalsi census survey data, which the Institute publishes at different aggregation levels \citep{Invalsi_IS}, the most informative one being LAU units or municipalities. 
We focus on the second year of high school ($10$-th school grade), being the last year of the compulsory education cycle for which Invalsi tests are designed (the last year of high school is beyond the compulsory education cycle). The subjects for which the test is designed for the school grade in scope are Italian and Mathematics.

We explore the association of Invalsi scores with the infrastructural state of municipalities in terms of their centrality degree expressed with the inner areas taxonomy, elaborated by the Italian National Institute of Statistics \citep[ISTAT hereinafter, ][]{InnerAreas}, availability of ultra-broadband internet connection in schools, and school accessibility using urban public transport. Geographical information is taken into account introducing a spatially structured latent effect in the regression model, defined at a higher aggregation level than municipalities, either provinces or catchment areas of infrastructural poles. Based on the prior belief that, besides the effect of explanatory variables, Invalsi scores tend to be closer in value across nearby areas than among ones far apart \citep{CAR}, we assume an Intrinsic Conditional Autoregressive structure \citep[hereinafter ICAR, ][]{BYM}. Since the scores in two subjects are available, a bivariate ICAR \citep{Mardia} spatial effect is modelled.

We attempt to ensure that covariates and spatial effects do not compete in explaining the target variable. This phenomenon is known in the literature as spatial confounding \citep{Hodges} and has been dealt with through several different methodologies \citep{Urdangarin23}, one of the latest and most promising ones being the Spatial+ approach \citep{Dupont}, which consists in removing the spatial variation from the covariates and regressing the response variable on the nonspatial component of covariates. We employ a recently proposed variant of the Spatial+ approach \citep{Urdangarin24} allowing to overcome the need to define a spatial model on covariates by leveraging on the spectral properties of the neighbouring structure of the data.

The analysis proposed throughout this paper follows a Bayesian paradigm and the main object of inference are therefore the marginal posterior distributions of both covariates and latent spatial effects. Due to the complexity of deriving the posterior marginals of interest, we resort to a deterministic approximation: the Integrated Nested Laplace Approximation \citep[INLA hereinafter, ][]{INLA, VBINLA}, a computational method implemented in a dedicated \texttt{R} environment \citep[\texttt{R-INLA},][]{INLAbook, Wang}, available at (\url{https://www.r-inla.org/home}). Since its introduction in 2009, one of the main fields of application of INLA has been spatial statistics indeed \citep{INLArev, Blangiardo}. In particular, multivariate spatial modelling of areal data is implemented in the \texttt{INLAMSM} package \citep{INLAMSM, INLAMSM2}, available \href{https://github.com/becarioprecario/INLAMSM}{on GitHub}, which we employ here for model fitting.
 
The remainder of this paper is structured as follows. In Section \ref{section:data} the data employed and the spatial structure referred to are described. In Section \ref{section:Model_outline} we outline the regression model used and the method followed to deal with spatial confounding. In Section \ref{section:Method} we summarise the application of the INLA and compare some possible model formulations. In Section \ref{section:results} we discuss the results of the models implemented.

\section{Student outcome data and infrastructural endowement} \label{section:data}
In this Section, the data on high school student outcomes together with some auxiliary data on the state of the infrastructure of Italian municipalities are described. 
Notice that Italian municipalities for which all relevant data are available amount to $873$ over $7904$ for the school year 2022/2023, the last school year for which all data are available on October 29, 2024. Data are obtained through the \texttt{SchoolDataIT} package \cite{SchoolDataIT}. 
 
\subsection{Invalsi scores}\label{Par:Invalsi} 
\begin{figure}
  \centering
  \includegraphics[width = 0.9\textwidth]{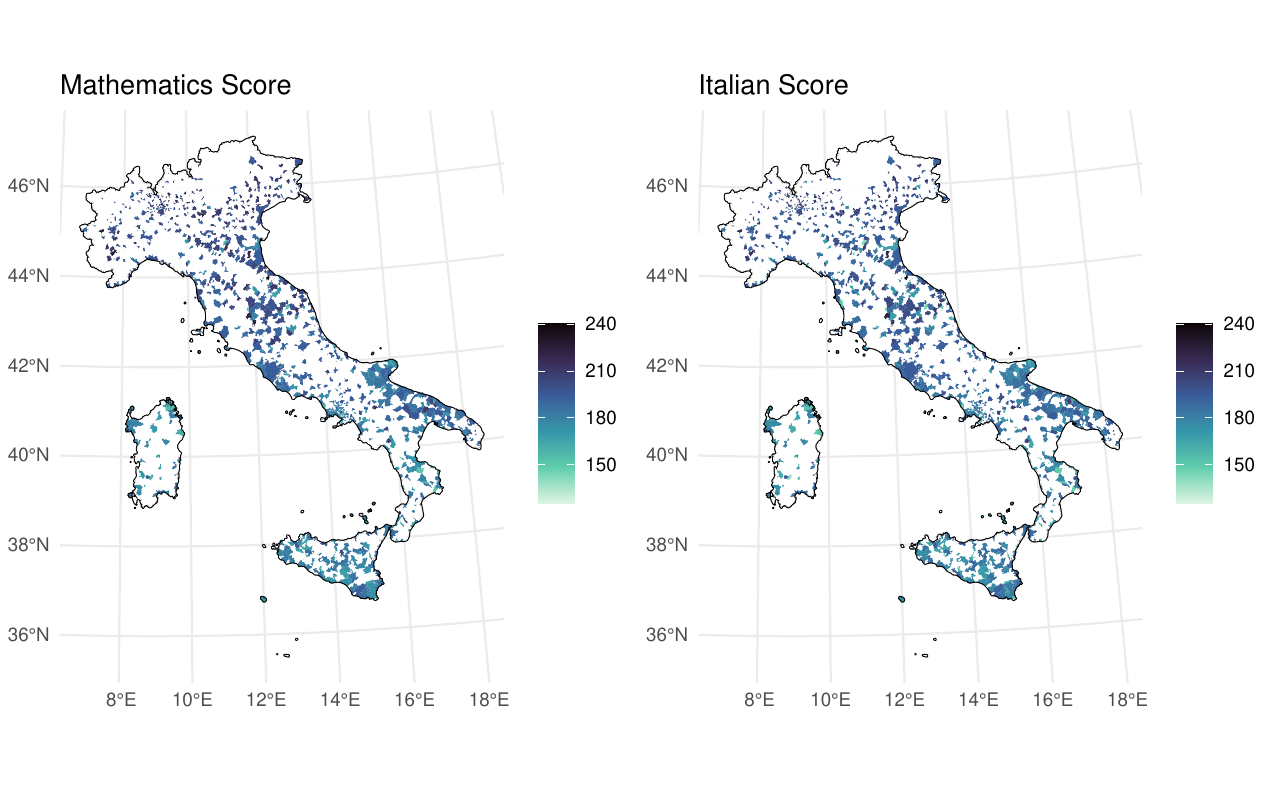} 
  \caption{Invalsi scores in 2022/2023, 2nd year of high school}
  \label{fig:Invalsi}
\end{figure}

The Italian Institute for the Evaluation of the Education System (INVALSI) is a research institute whose primary mission is to assess the quality of the public education system through a survey dedicated to the second and the last years of primary school, the last year of middle school, the second and the last years of high school. The Institute publishes both a sample survey including anonymized individual observations, and a census survey including the averaged scores at the region, province, and municipality levels. For privacy reasons, INVALSI only publishes the scores of the $978$ municipalities with at least two high schools \cite{Invalsi_IS}. Scores are expressed as the outcome of a psychometric model explaining the probability $\pi_{sq}$ that student $s$ passes the question $q$ (the same for all students) through two variables: student ability $b_s$ and item difficulty $d_q$ with such a formulation \citep[][Chapter 5]{PISA}:
 $$
 \pi_{sq} = \frac{e^{b_s - d_q}}{1 + e^{b_s - d_q}}
 $$
The ability estimator $b_s$ is then scaled to have a global mean of 200 points and a global between-students standard deviation of 40 points.
Invalsi scores in Mathematics and Italian at the 2nd year of high school in the school year 2022/2023 are displayed in Figure \ref{fig:Invalsi}. This and all other figures have been made using the \texttt{ggplot2 R} package \citep{ggplot}. Trentino-Alto Adige, a region with autonomous status in Italy, is absent due to the lack of availability of auxiliary variables.

\subsection{Auxiliary variables}  \label{Covariates}
Auxiliary information considered herein has been selected to synthesize the general infrastructural state of municipalities and the accessibility to schools. To the state of our findings, the most informative variables are the following:
\paragraph{Urban public transport} The municipality-level percentage of high schools located within 250 meters from a public urban transport hub, as reported in the School Buildings Section of the Unique School Data Portal \citep{MIUR}. Data are available for each public School building in Italy, except for the Trentino - Alto Adige region.  
 
\paragraph{Ultra-Broadband activation status} The municipality-level percentage of high schools where ultra-broadband connection had been implemented before September 1st, 2022. Ultra-broadband is defined as an internet connection with a maximum speed of 1 gigabit per second and a minimum guaranteed speed of 100 megabits/second until the peering, and open data regarding the implementation status are provided by \cite{BB}. Since the implementation plan does not regard all schools in Italy, the activation status is imputed to zero (not implemented) for all schools not listed in the Plan.

\paragraph{Inner Areas}\label{par:inner} The taxonomy of inner areas is published by the Italian National Institute of Statistics \citep{InnerAreas} and includes six classes; classes A and B identify the poles and the intermunicipality poles, namely the municipalities (A) or the clusters of contiguous municipalities (B) endowed with fundamental school, transport and health infrastructure. Classes from C to F are ranked in terms of travel time to the closest pole, referred to as the destination pole. 
Municipalities in classes A and B, namely the ones serving as destination poles, are labelled as central. Municipalities in classes C-D and E-F are labelled as intermediate and peripheral respectively. Dummy variables "Central" and "Peripheral" are defined according to this distinction.

\subsection{Spatial structure} \label{par:graph}
Considering $873$ municipalities over $7904$ leads to a sparse pattern of observational units. For the forthcoming analysis, it is thus convenient to define a less sparse spatial structure at the higher spatial aggregation level of macro-areas (see below). Say the total observational length is $N = 873$ municipalities, the number of macro-areas is $n$, and the number of municipalities within the $i$-th macro-area is $N_i$, then $N = \sum_{i=1}^{n} N_i$.

Two alternative definitions of the macro-areas are used, corresponding to two spatial models. The first level are provinces (NUTS-3 units), amounting to $n = 105$ macro-areas. Alternatively, infrastructural catchment areas are considered, defined as the ensemble of an infrastructural pole and all the intermediate and peripheral municipalities for which that pole is the destination pole. Infrastructural catchment areas amount to $n = 206$ units.

Macro-areas can be treated as the nodes of a graph $\mathcal{G}$ with $G = 3$ connected components, namely the continent and the islands of Sicily and Sardinia. The neighbourhood structure of $\mathcal{G}$ is described by the proximity matrix $\mathbf{W}$ whose generic element $w_{ij}$ is given by
$$
w_{ij} = \left \{ \begin{array}{l}
1 \iff i \sim j \\ 0 \, \, \, \, \text{otherwise}
\end{array} \right.
$$
where the $\sim$ sign denotes that two areas are neighbours.

\subsection{Spatial exploratory analysis of explanatory variables} \label{par:X}

\begin{figure}[htbp]
    \centering
    \includegraphics[width=0.9\textwidth]{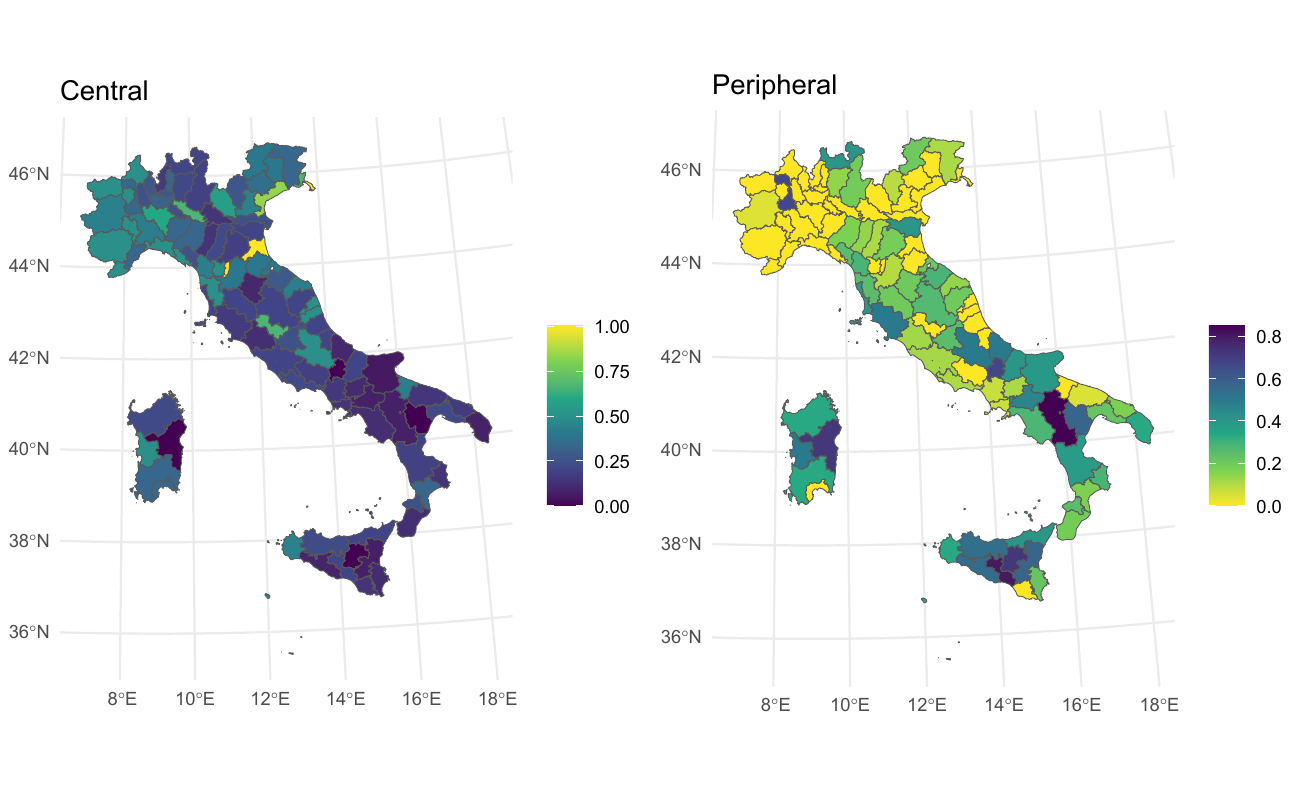} \\
    \includegraphics[width=0.9\textwidth]{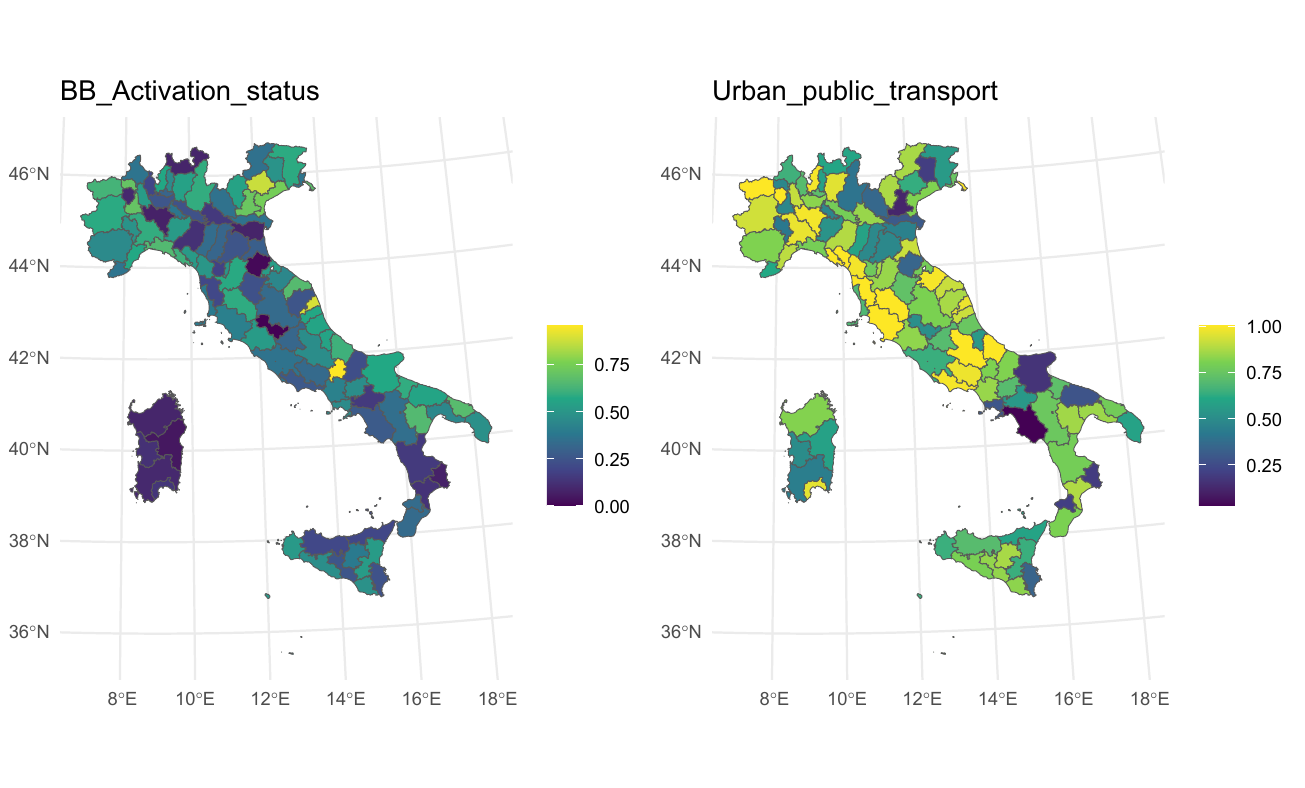} 
    \caption{Upper panel: proportion of central (left) and peripheral (right) municipalities per province.
    Lower panel: ultra-broadband availability (left) and urban transport accessibility (right) per province.}
    \label{fig:Xprov}
\end{figure}

In this Section, the spatial structure of explanatory variables at the province level is briefly explored.
In Figure \ref{fig:Xprov} auxiliary variables are mapped from municipalities to provinces by unweighted averages, i.e. the proportion of central and peripheral municipalities per province, and the unweighted averages of municipality-level proportions of schools served by ultra-broadband and urban transport are computed. Large-scale spatial variation is particularly evident in the first two variables, showing a higher concentration of infrastructural poles in the north and, vice-versa, a higher concentration of peripheral municipalities in the South. Mainly in the North, we also notice that some provinces have no peripheral municipalities at all, i.e. all non-central municipalities have a road travel time shorter than $41$ minutes \citep{InnerAreas} from the closest pole.

Concerning the ultra-broadband activation status, it is possible to observe a slight disadvantage in the mountainous inland regions and a strong disadvantage in the Sardinia region. The availability of public transport hubs shows a weak advantage for Central and Northwestern Italy.
In Table \ref{tab:MoranProv} the Moran's $I$ values is computed across provinces for the covariates. The standardised index $I_\mathrm{std}$ is obtained assuming the values $-1/104$ and $0.00459$ for the mean and the variance under the null hypothesis of no spatial autocorrelation \citep{Cliff_Ord}. For the first three variables the values of $I_\mathrm{std}$, suggesting a strong spatial autocorrelation, while the evidence of autocorrelation is weaker for the percentage of schools served by urban public transport.
The values of $I$ and $I_\mathrm{std}$ are computed with the \texttt{spdep} \texttt{R} package \citep{spdep}. 

%

\begin{table}[ht]
\centering
\begin{tabular}{lrr}
Variable & $I$ & $I_\mathrm{std}$  \\ \hline
Central & 0.2705 & 4.1338   \\  
Peripheral & 0.4236 & 6.3447   \\
Broadband avail. & 0.1908 & 2.9234 \\ 
Urban transport & 0.0662 & 1.1072  \\  
\hline
\end{tabular}
\caption{Moran's $I$ and standardised $I$ values for province-level averages of auxiliary variables.}
\label{tab:MoranProv}
\end{table}




\section{A bivariate spatial model for student scores}\label{section:Model_outline}

In this Section, the general features of the bivariate spatial model for the Invalsi scores in Mathematics and Italian are outlined and general notation is introduced.
Assume the Invalsi scores in Mathematics and Italian $y = (y_1^{\top}, \, y_2^{\top})^{\top}$ are defined as a vector of length $2N$ and modelled as follows:\\
\begin{equation}
y =  \mathbf{\tilde{X}}  \, \beta \,  + 
\tilde{\mathbf{\xi}} \, \mathbf{C} \, \beta_C \,  +
\, \tilde{\mathbf{\xi}} \,z \, + \, \varepsilon
\label{eq:model}
\end{equation}
\\
where $\mathbf{\tilde{X}} = I_2 \otimes \mathbf{X}$ and $\tilde{\xi} = I_2 \otimes \xi$. $I_2$ is the identity matrix of order $2$, $\mathbf{X}$ is the $N \times 4$ matrix of auxiliary variables (see Section \ref{Covariates}), and $\beta$ is a vector of fixed effects of length $8$. The $N \times n$ matrix $\xi$ is binary and maps the $n$ macro-areas onto $N$ municipalities. The $n \times 3$ matrix $\mathbf{C}$ is also binary and denotes which connected component (continent, Sicilia, Sardinia) each macro-area belongs to, $\beta_C$ is the vector of component-specific intercepts of length $6$. The bivariate latent spatial field $z = (z_1^{\top}, \, z_2^{\top})^{\top}$ is defined at the macro-area level and accounts for both the spatial variation and the correlation between Mathematics and Italian scores. Finally $\varepsilon = (\varepsilon_1^{\top} , \, \varepsilon_2^{\top})^{\top}$ is the random error vector distributed as follows:

\begin{equation}
\left\{
\begin{array}{ll}
\varepsilon_{1} \mid \omega_1 \overset{\text{iid}}{\sim}
N(\mathbf{0}, \omega_1)\\
\varepsilon_{2} \mid \omega_2, \alpha 
\overset{\text{iid}}{\sim} SN(m_{\omega_2, \alpha}, s_{\omega_2, \alpha}, \alpha)  
\end{array}
\right.
\label{eq:errors}
\end{equation} \\
$SN(\cdot)$ in \ref{eq:errors} denotes the Skew-Normal distribution \citep{SN} with location and scale parameters $m_{\alpha, \omega_2}$ and $ s_{\alpha, \omega_2}$ defined to ensure that $\mathbb{E}[\varepsilon_2 | \alpha] = 0 $ and $VAR[\varepsilon_2 | \alpha] = \omega_2$, and $\alpha$ is the shape parameter. This choice is due to the negative skewness in municipality-level Italian scores that neither auxiliary variables or spatial effects can explain (see Fig. \ref{fig:kernel} below). For interpretation reasons, in the remainder of this paper, we consider a transformation of $\alpha$, namely the skewness parameter $\gamma_1$, that has the property of lying approximately in the interval $]-1, 1[$:
$$
\gamma_{1} := \frac{4-\pi}{2}
\left(  \frac{2 \alpha^2}{\pi(1 + \alpha^2)} \right)^{\frac{3}{2}}
\left(  1 - \frac{2 \alpha^2}{\pi(1 + \alpha^2)} \right)^{-\frac{3}{2}}
$$
Following \cite{SNprior}, $\alpha$ is assigned a Penalised Complexity prior \cite{PCprior} with a given rate parameter $\lambda$. 
$\lambda = 4$ is chosen based on empirical considerations, i.e. balancing model complexity and fit. However, the posterior distribution of the skewness parameter does not appear to be sensitive to the choice of $\lambda$ \citep{SNprior}.\\

Covariate effects $\beta$ have $N(0, 10^3)$ non-informative priors, while priors for intercepts in $\beta_C$ are set as $N(180, 10^3)$, according to the expected global mean of Invalsi ability scores (Section \ref{Par:Invalsi}).
Precision parameters for the error terms, namely $\omega_{1}$ and $\omega_2$, have independent Gamma vague priors with shape $10^{-3}$ and rate $10^{-3}$.

\subsection{Modelling the spatial component} \label{par:ICAR}
Considering the neighbourhood structure outlined in Section \ref{par:graph}, $z$ is modelled as a bivariate ICAR defined on the graph $\mathcal{G}$. For a generic $i$-th node, with $i \in [1, n]$, the core assumption is that $z_i = (z_{i,1}, z_{i,2})^\top$ depends on neighbouring areas as follows:
\begin{equation}
z_{i} | z_{-i}, \Lambda \sim N \left(\sum_{j \sim i} \frac{w_{ij}}{d_i} z_{j}, \, \frac{1}{d_i} \Lambda^{-1}\right)
\label{eq:ICAR_local}
\end{equation}
where  $d_i := \sum_{s=1}^n w_{is}$ is the number of neighbours of node $i$, and $\Lambda$ is a matrix-valued global precision parameter. This representation is a special case of the model developed by \citep[][theorem 2.1, corollary 2]{Mardia} and implies a joint Normal prior on $z$ with zero mean and precision $\Lambda \otimes \mathbf{R}$,

%
where $\mathbf{R} := \mathbf{D} - \mathbf{W}$ is the Laplacian matrix of the graph $\mathcal{G}$ with $3$ connected components and $\mathbf{D} = \mathrm{diag}(d_1, d_2 \ldots d_n)$ is the degree matrix of $\mathcal{G}$.
Since $\mathbf{R}$ is singular with rank deficiency $3$ and $\pi(z|\Lambda)$ is therefore improper \citep{Hodges2003}, it is necessary to constrain $z$ to sum to zero within each connected graph component \citep{ICAR}. 
This is the reason for adopting component-specific intercepts $\beta_C$ in equation \ref{eq:model}.

To ease the interpretation of $\Lambda$ as the precision parameter of $z$, it is possible to cleanse it from the effect of the neighbourhood structure by defining a scaled version of $\mathbf{R}$ \cite{Sorbye} and reparametrising the precision of $z$ accordingly.
Since $\mathcal{G}$ is disconnected, each component-specific block of $\mathbf{R}$ is multiplied by the relevant typical variance, namely the geometric mean of the diagonal of the corresponding block of its pseudoinverse, following the methodology proposed by \cite{Freni}. It is therefore possible to define a precision parameter $\Lambda_\mathrm{scaled}$ which is not confounded with graph-induced effects. The scaled precision is assigned a Wishart prior \citep{Gelman} with $2k+1$ degrees of freedom and scale parameter equal to the identity matrix, i.e. $\Lambda_\mathrm{scaled} \sim \mathrm{Wishart}_{k}( I_k, 2k+1)$, with $k=2$ \citep{INLAMSM}.



\subsection{Spatial confounding}
Spatial confounding can occur when a regression model includes a spatially structured latent random variable correlated with some explanatory variables. This issue implies a competition between $\mathbf{X}$ and $z$ in explaining $y$, introducing bias in the estimation of $\beta$. Several approaches have been developed in almost two decades of literature \citep{Urdangarin23, DupontArXiv}, starting from the intuitive solution of constraining spatial random effects to be linearly independent of covariates \citep{RHZ, Hodges}, which goes under the name of restricted spatial regression (RSR hereinafter). This is done by projecting random effects onto the subspace orthogonal to the design matrix $\mathbf{X_\mathrm{tot}} = (\xi\mathbf{C}, \mathbf{X})$, i.e. conditioning $z$ to the constraint $\mathbf{X_\mathrm{tot}}(\mathbf{X_\mathrm{tot}}^{\top}\mathbf{X_\mathrm{tot}})^{-1}\mathbf{X_\mathrm{tot}}^{\top} \, \xi z = \mathbf{0}$. The idea behind RSR is to rule out confounding bias when estimating the effects of covariates. RSR can be extended to the multilevel case \citep{Nobre}, where spatial effects are defined at a higher scale than observations, as we do in the present framework. Nevertheless, under RSR the posterior means of covariate effects tend to approximate those of the nonspatial model \citep{Khan}, which leads to biased estimates of $\beta$ by ignoring the presence of a latent spatial field and assuming independent errors \citep{DupontArXiv}.
This suggests exploring further methodologies to deal with spatial confounding.

The Spatial+ approach \citep{Dupont} involves the adjustment of covariates, instead of constraining spatial effects, by decomposing the former as the sum of a spatial and a nonspatial component. Here a variant of the Spatial+ method, developed by \cite{Urdangarin24}, is applied while working in a multilevel framework. This innovative methodology consists in extracting the spatial component of covariates without requiring an explicit spatial model on $\mathbf{X}$. Following \cite{Lamouroux}, we label this methodology as Spatial+ 2.0. 

In our multilevel framework, the value of the $m$-th covariate $\mathbf{X_{\cdot m}}$ observed in municipality $h$ belonging to macro-area $i$ can be decomposed as
$$
x_{ih;m} = \bar{x}_{i;m} + \Delta x_{ih;m}
$$
being $\bar{x}_{i;m}$ the unweighted average value of the covariate within the $i$-th macro-area; the term $\Delta x_{ih;m}$ represents municipality-level noise. In matrix form, this decomposition is: $\mathbf{X}= \xi \bar{\mathbf{X}} + \mathbf{\Delta X}$, where $\bar{\mathbf{X}} = (\xi ^{\top} \xi)^{-1} \xi^{\top} \mathbf{X}$.

Consider the eigendecomposition of the Laplacian matrix \citep{Urdangarin24}: 
$$
\mathbf{R} = \mathbf{VLV}^{\top}
$$
where the eigenvalues in $\mathbf{L}$ are in decreasing order and the eigenvectors in $\mathbf{V}$ have a decreasing number of oscillations. 

Within a generic component of the connected graph, the eigenvector associated with the lowest non-null eigenvalue follows a linear spatial trend (i.e. the Fiedler vector of the relevant subgraph), the one related to the second non-null eigenvalue follows a quadratic trend (one oscillation), and so on.
For an appropriately chosen $n \times 4$ matrix $\mathbf{b}$, $\mathbf{\bar{X}}$ can be expressed as a linear combination of $\mathbf{V}$:
$$
\mathbf{\bar{X}} = \mathbf{Vb}
$$
Intuitively, the spatial component of $\mathbf{\bar{X}}$ is determined by the last columns of $\mathbf{V}$ \citep{Urdangarin24}: without loss of generality, $\mathbf{\bar{X}}$ is decomposed into:
$$
\mathbf{\bar{X} }= \mathbf{\bar{X}^{(NS)}} + \mathbf{\bar{X}^{(S)}} +\mathbf{ \bar{X}^{(0)}}
$$
where $\mathbf{\bar{X}^{(NS)}}$ is the nonspatial component, given by the linear combination of the first $n-G-K$ eigenvectors (with $G=3$ connected graph components), $\mathbf{\bar{X}^{(s)}}$ is the combination of the eigenvectors associated with the last $K$ nonzero eigenvalues and represents the spatial component, and $\mathbf{\bar{X}^{(0)}}$ is the combination of the $G=3$ eigenvectors in the null space of the Laplacian matrix, constant within each connected component. 
To remove spatial confounding, it is sufficient to consider $\xi \left(\mathbf{\bar{X}^{(NS)}} \right) + \mathbf{ \Delta X}$ as the covariate matrix in the regression model. 



\begin{figure}
  \centering
  \includegraphics[width=0.9\textwidth]{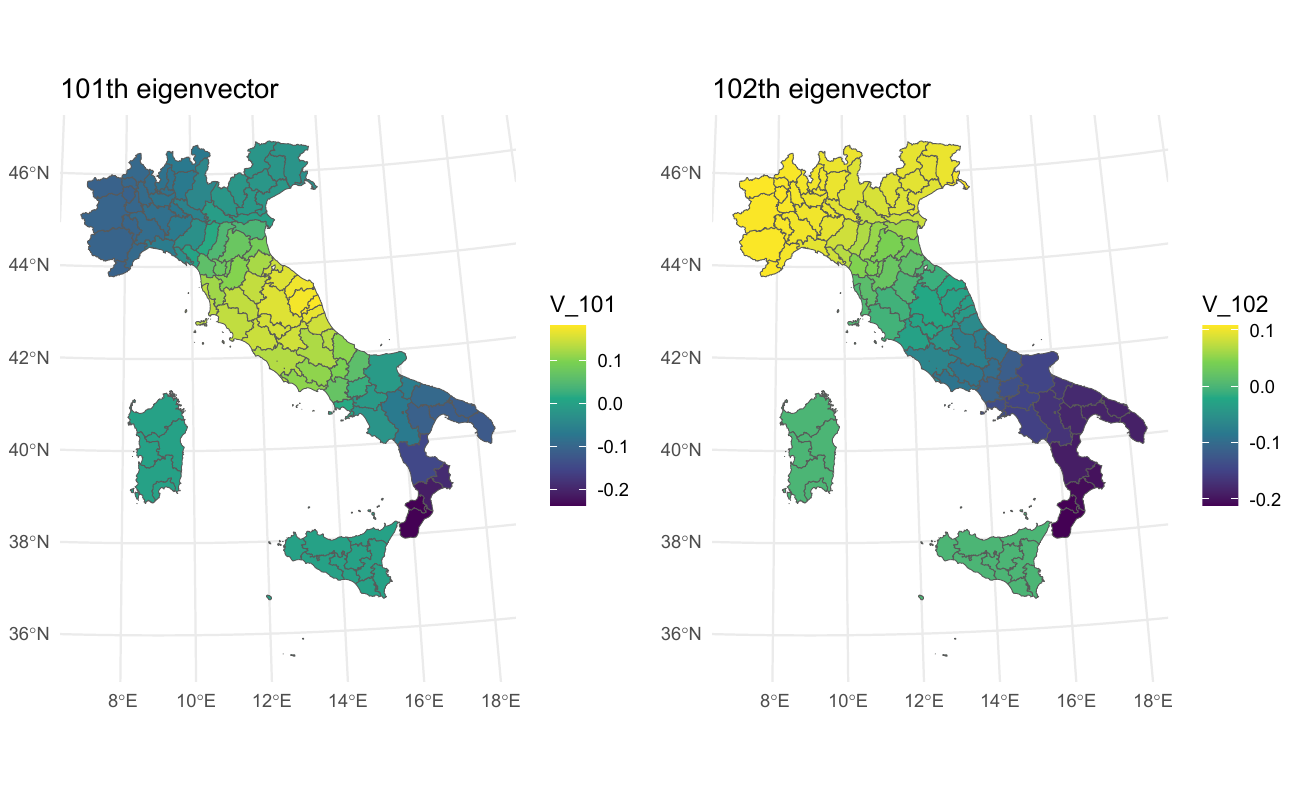} 
  \caption{Lowest frequency eigenvectors, provinces}
  \label{fig:eigen_prov}
\end{figure}

In Figure \ref{fig:eigen_prov} the eigenvectors corresponding to the last two non-zero eigenvalues of $\mathbf{R}$ at the macro-area level of provinces for the continent graph component are plotted. It is possible to see that the second-last eigenvector follows a quadratic trend with one oscillation, whereas the last eigenvector follows a linear North-South trend.  

\subsubsection{Identification of the spatial variation in the covariates}\label{par:k}
The Spatial+ procedure requires the choice of the number $K$ of eigenvectors to define $\mathbf{\bar{X}^{(NS)}}$. A documented approach \citep{Urdangarin24, Lamouroux} is to choose the value of $K$ minimizing the Watanabe-Akaike Information Criterion \citep[WAIC,][]{WAIC}. Based on this method, our first strategy is searching for the optimal number of eigenvectors to be removed for each explanatory variable, subject to the following constraints.

When $z$ is defined at the province level (91 areas in the continent, 9 in Sicily, 5 in Sardinia), we remove a maximum of 9 eigenvectors in the continent, and 1 for each of the two islands, whereas when $z$ is defined at the level of infrastructural catchment areas (186 macro-areas in the continent, 13 in Sicily, 7 in Sardinia) we remove up to 18 eigenvectors from the continent, 2 from Sicily and 1 from Sardinia.

As an alternative strategy, we removed the smallest number of eigenvectors for $\mathbf{\bar{X}^{(NS)}}$ not to display evidence of autocorrelation according to Moran's $I$ index. 
For the first three variables at the province level, removing the last $4$ eigenvectors leads to small standardized $I$ values ($0.79697$, $1.3854$, $1.2397$), suggesting that spatial structure in these variables is driven by a linear trend over the continent, as seen in Figure \ref{fig:Xprov}. For infrastructural catchment areas, doing the same thing with the proportion of central and peripheral municipalities would lead to standardized $I$ values of $1.2522$ and $0.4598$ respectively; again, this can be interpreted as the presence of a linear trend. For the last two covariates, instead, it is necessary to remove a higher amount of eigenvectors, which suggests the presence of spatial variation on a smaller scale before accepting the hypothesis of no spatial autocorrelation. Details on the removal patterns are in Appendix \ref{Appendix:A}.

\section{Model estimation using R-INLA}\label{section:Method}

Keeping in mind the model described in equation \ref{eq:model}, on a first stance it is convenient to define the sets of latent variables $\theta$ and hyperparameters $\Psi$:
$$
\theta :=  \lbrace \beta_C,  \beta, z  \rbrace  \quad \text{and} \quad
\Psi :=  \lbrace \Lambda_{\mathrm{scaled}}, \omega_1 , \omega_2, \alpha \rbrace
$$

Our aim is inference on posterior marginals $\pi(\theta_i | y)$ and $\pi(\Psi_l | y)$, with $i=1,\ldots, 14 + 2n$ and $l = 1,\ldots, 6$ which considering the model described in Section \ref{section:Model_outline} cannot, however, be obtained in closed form. Provided that model assumptions imply a joint Gaussian prior on $\theta$, this issue can be overcome by approximating the posterior distributions of interest through the INLA method \citep{INLA}. For the model and the data at hand, two alternative numerical approximation methods are considered: the Simplified Laplace \citep{INLA} and the more recently developed Variational Bayes \citep{VB} approximation (respectively SL and VB hereinafter).
%
Posterior marginals are then computed by numerical integration of $\pi(\theta_i, \Psi | y)$ and $\pi(\Psi | y)$ over a set of values of $\Psi$; locating the posterior mode of $\Psi$ is crucial for all models discussed in the next session \ref{Par:results}.

In what follows, we focus mainly on the VB approximation, due to difficulties in locating the mode of $\Psi$ with the SL approximation. The comparison between models estimated with these two approaches is shown in Appendix \ref{Appendix:B}.

All calculations in this paper have been carried out using the \texttt{2024.10.13} version of \texttt{R-INLA}. In order to make results as reproducible as possible, \texttt{R-INLA} is run on a single core \citep{Wang} and internal optimization is disabled.



\subsection{Model assessment} \label{par:criteria}

In this Section, some alternative model formulations are compared using a set of selection criteria internally computed by \texttt{R-INLA}: the negative Log Pseudo Marginal Likelihood \citep[LPML,][]{Geisser, Gelfand}, i.e. minus the logarithmic sum of the Conditional Predictive Ordinates \citep{CPO, CPOINLA}, and the Watanabe-Akaike Information criterion \citep[WAIC,][]{WAIC}, following the formulation of \citet{GelmanWAIC}, the Deviance Information Criterion \citep[DIC,][]{DIC}, alongside with the Mean Squared Error of posterior predictive response averages.

Models defined with ICAR random effects are compared in Table \ref{tab:ICAR_diagnostics}. "Base" and "RSR" denote the model with no correction for spatial confounding and the RSR model respectively.
S+(1) and S+(2) are Spatial+2.0 models with province-level latent effects with two different eigenvector removal patterns: the former is the most conservative one for which no evidence for autocorrelation in the covariates is found (see Section \ref{par:X}), the latter is the one with smallest WAIC.
S+(3) and S+(4) are the Spatial+2.0 models developed with the same strategy but with latent effects defined at the level of infrastructural catchment areas. Detailed eigenvector removal patterns are shown in Appendix \ref{Appendix:A}.

Removing spatial autocorrelation from covariates based on Moran's test allows for some barely noticeable improvements in inference. Using finer support for the latent effects improves the fitting but this gain is outweighed by increased complexity (overall, the DIC increases).The model with province-level spatial effects is overall preferable based on all three metrics WAIC, DIC and LPML.
RSR appears to perform poorly in both cases if compared to the base model. Furthermore, posterior means of $\beta$ obtained by RSR result quite close to those of the nonspatial model, while credible intervals are narrower, which is consistent with the lesser coverage of Type-S errors in RSR models \citep{Khan}. 

In Appendix \ref{Appendix:B} the results of a broader set of models are shown, including models with no random effects, with unstructured random effects and with two independent ICAR random effects; a focus on the estimates of $\beta$ under the nonspatial model and under RSR is in Appendix \ref{Appendix:null_RSR}. 

Lastly, in Appendix \ref{Appendix:C} the results of a different model, the proper CAR \citep{PCAR_Gelfand}, are summarised. The core feature of this formulation is introducing an additional parameter to account for the strength of spatial association.

\begin{table}[ht]
\centering
\begin{tabular}{llrrrrrr}
  \toprule
  $z$ level & Model & -LPML & WAIC & DIC & MSE \\ 
  \midrule
  Prov & Base &6689.917 & 13379.149 & 13379.808 & 235.551 \\ 
  Prov & RSR  &6764.094 & 13526.761 & 13526.476 & 251.886 \\ 
  Prov & S+(1)&6689.672 & 13378.651 & 13379.311 & 235.326 \\ 
  Prov & S+(2)&\textbf{6689.500} & \textbf{13378.303} & \textbf{13378.909} & 235.189 \\  \midrule 
  Pole & Base &6694.435 & 13387.776 & 13388.726 & 232.320 \\ 
  Pole & RSR  &6754.037 & 13505.502 & 13507.315 & 239.313 \\ 
  Pole & S+(3)&6694.443 & 13387.747 & 13388.711 & \textbf{231.811} \\ 
  Pole & S+(4)&6694.055 & 13387.018 & 13387.948 & 231.838 \\ 
   \botrule 
   
\end{tabular}
\caption{Model diagnostics for 8 ICAR model formulations: spatial aggregation level of $z$, spatial confounding treatment, negative Log Pseudo Marginal Likelihood, Watanabe-Akaike Information criterion, Deviance Information Criterion, Mean Squared Error of posterior predictive response averages.}
\label{tab:ICAR_diagnostics}
\end{table}
\section{Results}\label{Par:results} \label{section:results}

In Table \ref{tab:fix} the estimated effects of covariates for the province-level ICAR are resumed, under both the base formulation and the S+(2) modification. Boundaries of credible intervals correspond to the $5$-th and $95$-th percentiles. Covariates in the deconfounded model have been scaled to keep the same variance as in the base model.

\begin{table}[ht]
\centering
\begin{tabular}{ll|rrrr|rrrr}
  \toprule
  && \multicolumn{4}{c|}{Base model} & \multicolumn{4}{c}{S+(2)}\\
 & Subj & mean & sd & LB & UB & mean & sd & LB & UB \\ 
  \midrule
Continent & MAT & 191.399 & 0.961 & 189.515 & 193.285 & 193.332 & 0.855 & 191.656 & 195.009 \\ 
  Continent & ITA & 187.107 & 0.999 & 185.137 & 189.056 & 188.585 & 0.863 & 186.886 & 190.270 \\ 
  Sicily & MAT & 177.764 & 1.496 & 174.829 & 180.698 & 178.386 & 1.369 & 175.702 & 181.070 \\ 
  Sicily & ITA & 176.884 & 1.550 & 173.835 & 179.915 & 177.273 & 1.417 & 174.489 & 180.046 \\ 
  Sardinia & MAT & 174.325 & 2.197 & 170.017 & 178.634 & 174.561 & 2.159 & 170.327 & 178.795 \\ 
  Sardinia & ITA & 171.914 & 2.267 & 167.460 & 176.351 & 172.126 & 2.208 & 167.790 & 176.450 \\ 
  Central & MAT & 2.706 & 0.910 & 0.922 & 4.490 & 2.527 & 0.890 & 0.781 & 4.273 \\ 
  Central & ITA & 2.379 & 0.996 & 0.433 & 4.338 & 2.460 & 0.979 & 0.547 & 4.386 \\ 
  Peripheral & MAT & -2.200 & 1.005 & -4.171 & -0.228 & -2.018 & 0.958 & -3.897 & -0.139 \\ 
  Peripheral & ITA & -1.845 & 1.049 & -3.901 & 0.215 & -1.793 & 1.000 & -3.753 & 0.170 \\ 
  BB Activation & MAT & 3.331 & 1.074 & 1.226 & 5.437 & 3.262 & 1.049 & 1.205 & 5.319 \\ 
  BB Activation & ITA & 2.296 & 1.126 & 0.090 & 4.509 & 2.130 & 1.100 & -0.024 & 4.291 \\ 
  Urban transport & MAT & 2.466 & 1.043 & 0.420 & 4.513 & 2.501 & 1.044 & 0.453 & 4.549 \\ 
  Urban transport & ITA & 2.841 & 1.060 & 0.765 & 4.924 & 2.838 & 1.060 & 0.762 & 4.919 \\ 
   \botrule
\end{tabular}
\caption{Posterior summaries of intercepts and covariate effects when $z$ is defined as a province-level ICAR, under the base model and the Spatial+2.0 model (optimal combination of eigenvector removal under the WAIC metric)}
\label{tab:fix}
\end{table}

Modelling Italian scores appears to be generally subject to higher uncertainty. For both subjects, differences between the continent and the islands are strong: almost $15$ Invalsi points on average between the continent and Sicily, more than $15$ points between the continent and Sardinia, with non-overlapping credible intervals. Central municipalities have an expected advantage of $2.706$ points over intermediate municipalities in Mathematics test, while this expectation slightly falls to $2.527$ points once the share of infrastructural poles in each province is corrected with S+(2). The relative effect of central municipalities on Italian scores is comparable and slightly lower. The difference between intermediate and peripheral municipalities is lower in expected value and not even significant for Italian scores. A municipality in which all schools are provided with ultra-broadband connection has an expected advantage of more than $3$ Invalsi points in the Mathematics score over one in which the connection is completely lacking, the effect being weaker on Italian scores (and possibly not significant once the spatial structure is removed from the covariate). Lastly, the availability of urban public transport is associated with a significant advantage in Invalsi scores, since a municipality where all schools are reachable has an expected advantage of almost $2.5$ points in Mathematics scores and about $2.8$ points in Italian scores. 

\begin{figure}
  \centering
  \includegraphics[width=0.9\textwidth]{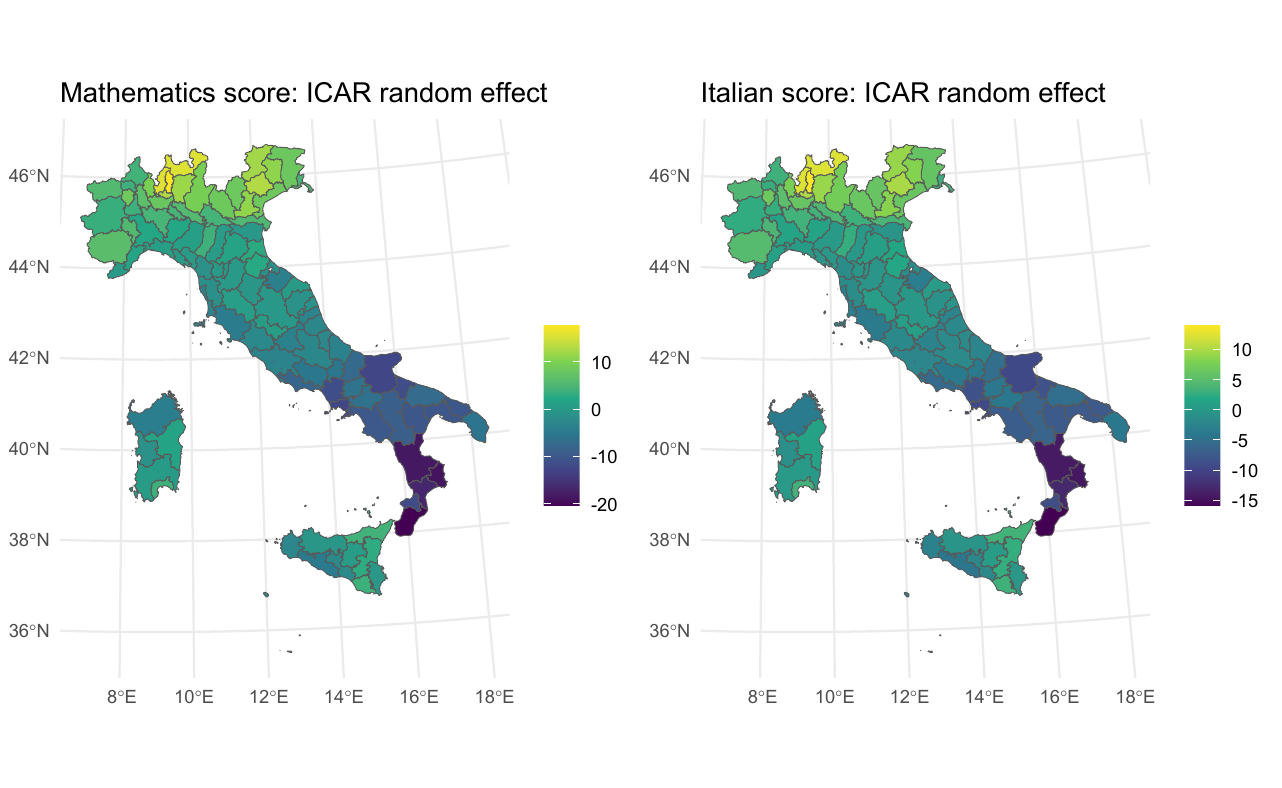} 
  \caption{Expected values of $z$ modelled as a province-level ICAR and applying S+(2) correction}
  \label{fig:zprov}
\end{figure}

In Figure \ref {fig:zprov} the expected spatial effect $\mathbb{E}[z|y]$ under the S+(2) model is plotted. 
Territorial gaps in Invalsi scores are severe, as one can argue from the range of $\mathbb{E}[z|y]$. Focusing on the continent, the Calabria region appears particularly vulnerable, while Lombardia turns out to be the most advantaged region. 

\begin{figure}
  \centering
  \includegraphics[width=0.9\textwidth]{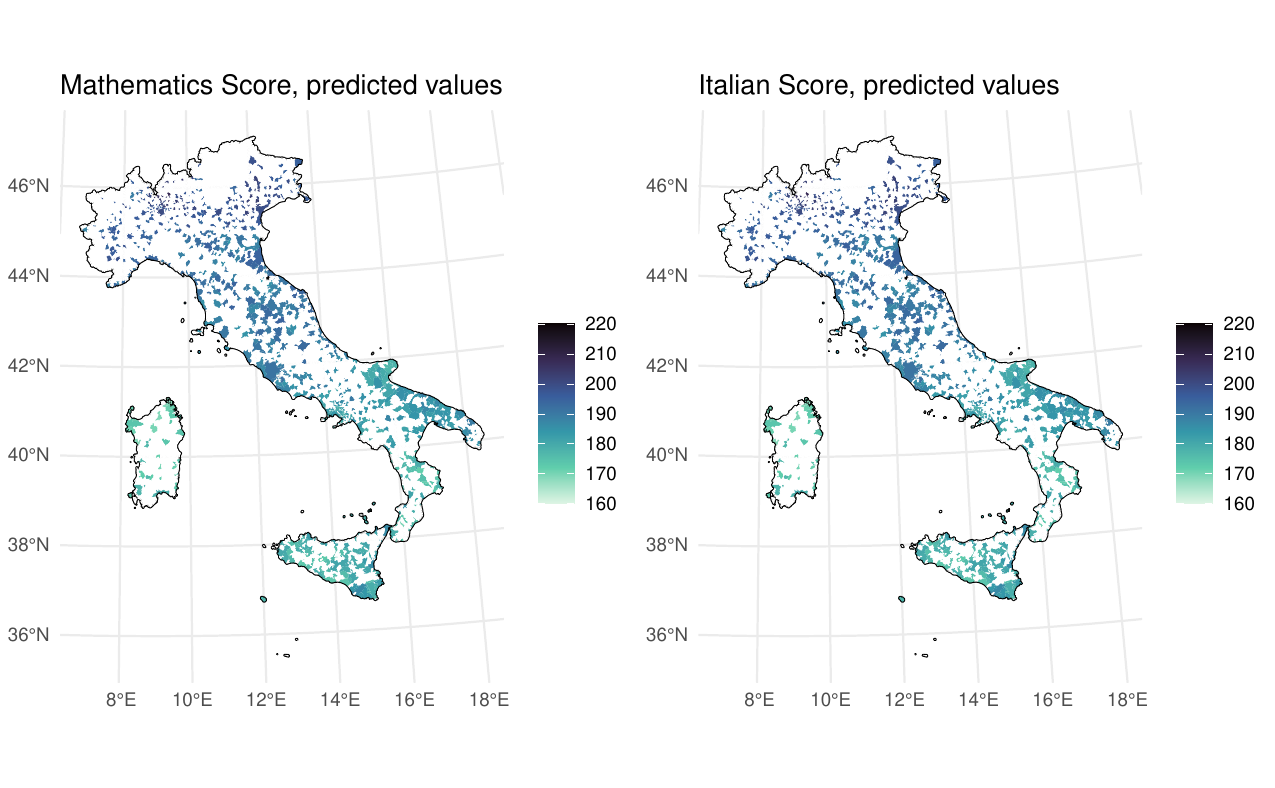} 
  \caption{Predicted values of Invalsi scores using as a province-level ICAR latent effect and applying the S+(2) correction}
  \label{fig:yhatmun}
\end{figure}

In Figure \ref{fig:yhatmun} the predicted values of $y$ are shown, using the same model. The model captures the spatial trend, but still leaves a high municipality-level noise unexplained, as the high error variances suggest ($\omega_1$ and $\omega_2$ in Table \ref{tab:hyperpar}). 

The highest scores are estimated in the municipalities of Lecco (Lombardia), with expected scores of $216.072$ points in Mathematics (observed score of $214.042$ points) and $207.637$ points in Italian (observed $204.990$ points) and Merate (province of Lecco), with expected scores of $216.133$ points in Mathematics (observed score $229.352$ points) and $207.274$ points in Italian (observed value $216.045$ points). 

Lowest scores in Mathematics are estimated in the municipalities of La Maddalena (province of Sassari, Sardinia) at $169.351$ points and Oppido Mamertina (province of Reggio Calabria) at $170.301$ points  (observed $165.814$ and $170.105$ points respectively). Lowest scores in Italian are estimated in the municipalities of La Maddalena at $170.016$ points  and Bosa (province of Oristano, Sardinia) at $169.787$ points (observed $168.071$ and $168.890$ points respectively).

\begin{figure}
  \centering
  \includegraphics[width=0.6\textwidth]{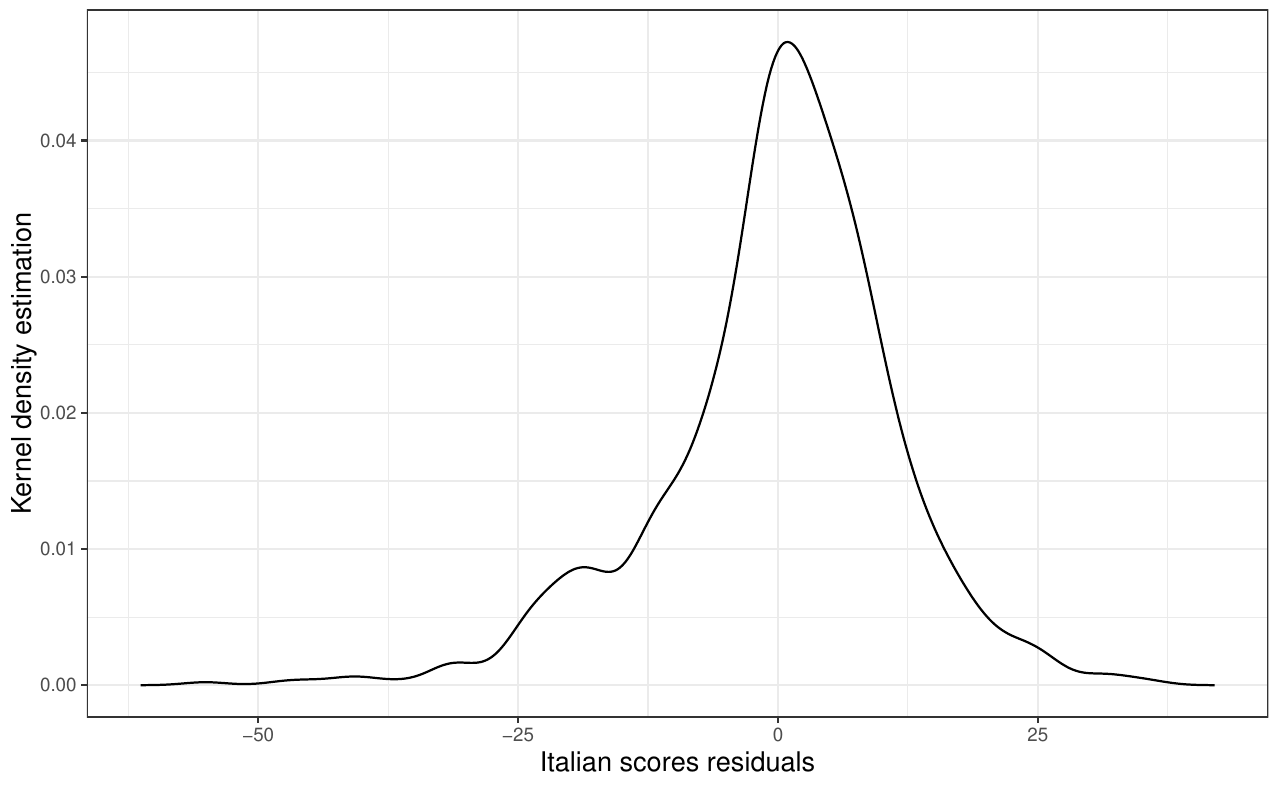} 
  \caption{Kernel density estimation of the residuals of Italian scores under the S+(2) model}
  \label{fig:kernel}
\end{figure}

Finally, in Table \ref{tab:hyperpar} the posterior summaries for hyperparameters $\Psi$ are displayed. Please notice that the precision of $z$ has been scaled, hence variances $\sigma_1^2$ and $\sigma_2^2$ do not depend on the graph-induced effect. The variance of spatial effects is higher in Mathematics scores ($\sigma_1^2$), while Italian scores have a higher amount of unexplained noise ($\omega_2$). Correlation between the two scores is taken into account through the correlation between the two ICAR fields $\rho$, which turns out to be high, consistent with Figure \ref{fig:zprov}.
Lastly, the choice of modelling Italian scores as a Skew-Normal variable is corroborated by the posterior distribution of $\gamma_1$, whose credible interval ranges far from zero. Kernel density estimation of residuals in Italian scores 
is shown in Figure \ref{fig:kernel}. Negative skewness is easily noticeable. Density is estimated by the Gaussian kernel, using the Silverman's thumb rule to define the bandwidth \citep[][Section 3.4.2]{Silverman}

\begin{table}[ht]
\centering
\begin{tabular}{ll|rrrr|rrrr}
  \toprule
  && \multicolumn{4}{c|}{Base model} & \multicolumn{4}{c}{S+(2)}\\
 & Subj & LB & Median & UB & sd & LB & Median & UB & sd \\ 
  \midrule
 $\sigma_1^2$ & MAT & 16.621 & 27.414 & 46.208 & 7.578 & 17.288 & 28.722 & 46.823 & 7.562 \\ 
 $\sigma_2^2$ & ITA  & 10.277 & 18.156 & 33.042 & 5.843 & 10.672 & 18.832 & 32.760 & 5.663 \\ 
 $\rho$ &     & 0.894 & 0.975 & 0.995 & 0.027 & 0.884 & 0.976 & 0.994 & 0.030 \\ 
  $\omega_1$ & MAT & 100.830 & 110.924 & 122.138 & 5.426 & 100.712 & 110.790 & 121.982 & 5.417 \\ 
  $\omega_2$ & ITA & 119.507 & 132.104 & 145.888 & 6.718 & 119.368 & 131.795 & 145.646 & 6.692 \\
  $\gamma_1$ & ITA & -0.493 & -0.371 & -0.232 & 0.067 & -0.493 & -0.369 & -0.232 & 0.066 \\ 
   \botrule
\end{tabular}
\caption{Posterior summaries of hyperparameters when $z$ is defined as a province-level ICAR, under the base model and the Spatial+2.0 model (optimal combination of eigenvector removal under the WAIC metric)}
\label{tab:hyperpar}
\end{table}

\section{Concluding remarks}

While a good body of literature studies student ability scores at the individual level, we have turned the analysis to a different framework, attempting to explain the geographical distribution of Invalsi scores based on infrastructural variables and using a multivariate spatial regression model. For a better understanding of the spatial variation, the precision of the spatial effect was scaled to account for discontinuities implied by the presence of two islands. The typical skew distribution in the Italian Invalsi scores was accounted for by relaxing the Normality assumption and fitting a Skew-Normal likelihood, with evidence for negative skewness. Lastly, to avoid explanatory variables being confounded with the spatial effect, we cleansed them from low-frequency spatial trends in estimating their effect on Invalsi scores.

The infrastructural state of schools and municipalities results in a significant impact on student performances, also when their effect is separated from spatial information. The classification of Italian municipalities into central, intermediate, and peripheral allows to report a noticeable advantage of the first to the second, while the difference between intermediate and peripheral municipalities is weaker. Our results do highlight the overall vulnerability of inner areas under the educational dimension.

Our analysis of Invalsi scores includes a spatially structured latent field defined at the level of macro-areas, either provinces or infrastructural catchment areas. We find that the between-macro-areas spatial effect is indeed a strong driver of Invalsi scores, in addition to between-municipalities factors. This confirms the strength of the territorial divides shaping many aspects of Italian society, especially the North-South gap.

Still keeping our focus on infrastructural explanatory variables, this analysis can undergo some possible developments: first and foremost it can be extended to different school grades and different years. The choice of statistical models, both in terms of likelihood and prior assumptions, could be extended as well, including more tailored models. The PCAR model, for instance, may represent a worthwhile extension, though questions like improving the interpretation of precision parameters remain open, further research being needful to this aim.

Overall, the mapping of infrastructure access and social vulnerability requires adequate statistical computational methods, and \texttt{R-INLA} appears a very flexible tool to this aim.

\section*{Acknowledgement}

\begin{itemize}
\item \textbf{Competing Interests.} The authors declare to have no competing interests. \\

\item \textbf{Funding.} Leonardo Cefalo and Alessio Pollice declare that this study was funded by the European Union - NextGenerationEU, Mission 4, Component 2, in the framework of the GRINS -Growing Resilient, INclusive and Sustainable project (GRINS PE00000018 – CUP H93C22000650001). The views and opinions expressed are solely those of the authors and do not necessarily reflect those of the European Union, nor can the European Union be held responsible for them. \\

\item \textbf{Preprint.} A preliminary version of this paper, named \textit{Multilevel Bivariate Areal Modelling for School Data: an application with R-INLA} was presented at the \textit{52nd Scientific Meeting of the Italian Statistical Society} (SIS2024) on June 19th, 2024, and published in Pollice A., Mariani P., Methodological and Applied Statistics and Demography, volume III, SIS 2024, Short Papers, Contributed Sessions 1, ISBN 978-3-031-64430-6.
\end{itemize}

%
%
\bibliography{references}

\begin{thebibliography}{}
\renewcommand{\doi}[1]{\url{https://doi.org/#1}}
\bibcommenthead

\bibitem [\protect \citeauthoryear {%
Agasisti%
\ \BBA {} Vittadini%
}{%
Agasisti%
\ \BBA {} Vittadini%
}{%
{\protect \APACyear {2012}}%
}]{%
Agasisti}
\APACinsertmetastar {%
Agasisti}%
\begin{APACrefauthors}%
Agasisti, T.%
\BCBT {}\ \BBA {} Vittadini, G.%
\end{APACrefauthors}%
\unskip\
\newblock
\APACrefYearMonthDay{2012}{}{}.
\newblock
{\BBOQ}\APACrefatitle {Regional economic disparities as determinants of
  student's achievement in {Italy}} {Regional economic disparities as
  determinants of student's achievement in {Italy}}.{\BBCQ}
\newblock
\APACjournalVolNumPages{Research in Applied Economics}{4}{2}{33,}
\newblock
\begin{APACrefDOI} \doi{http://dx.doi.org/10.5296/rae.v4i2.1316}
  \end{APACrefDOI}
\newblock

\newblock

\PrintBackRefs{\CurrentBib}

\bibitem [\protect \citeauthoryear {%
Azzalini%
\ \BBA {} Capitanio%
}{%
Azzalini%
\ \BBA {} Capitanio%
}{%
{\protect \APACyear {2014}}%
}]{%
SN}
\APACinsertmetastar {%
SN}%
\begin{APACrefauthors}%
Azzalini, A.%
\BCBT {}\ \BBA {} Capitanio, A.%
\end{APACrefauthors}%
\unskip\
\newblock
\APACrefYear{2014}.
\newblock
\APACrefbtitle {The Skew-Normal and Related Families} {The skew-normal and
  related families}\ (\BVOL~3).
\newblock
\APACaddressPublisher{}{Cambridge University Press}.
\PrintBackRefs{\CurrentBib}

\bibitem [\protect \citeauthoryear {%
Bakka%
\ \protect \BOthers {.}}{%
Bakka%
\ \protect \BOthers {.}}{%
{\protect \APACyear {2018}}%
}]{%
INLArev}
\APACinsertmetastar {%
INLArev}%
\begin{APACrefauthors}%
Bakka, H.%
, Rue, H.%
, Fuglstad, G\BHBI A.%
, Riebler, A.%
, Bolin, D.%
, Illian, J.%
\BDBL {}Lindgren, F.%
\end{APACrefauthors}%
\unskip\
\newblock
\APACrefYearMonthDay{2018}{}{}.
\newblock
{\BBOQ}\APACrefatitle {Spatial modeling with {R-INLA}: A review} {Spatial
  modeling with {R-INLA}: A review}.{\BBCQ}
\newblock
\APACjournalVolNumPages{WIREs Computational Statistics}{10}{6}{e1443,}
\newblock
\begin{APACrefDOI} \doi{https://doi.org/10.1002/wics.1443} \end{APACrefDOI}
\newblock

\newblock

\PrintBackRefs{\CurrentBib}

\bibitem [\protect \citeauthoryear {%
Barrett%
, Treves%
, Shmis%
\BCBL {}\ \BBA {} Ambasz%
}{%
Barrett%
\ \protect \BOthers {.}}{%
{\protect \APACyear {2019}}%
}]{%
WB}
\APACinsertmetastar {%
WB}%
\begin{APACrefauthors}%
Barrett, P.%
, Treves, A.%
, Shmis, T.%
\BCBL {} Ambasz, D.%
\end{APACrefauthors}%
\unskip\
\newblock
\APACrefYearMonthDay{2019}{}{}.
\newblock
\APACrefbtitle {The impact of school infrastructure on learning: A synthesis of
  the evidence.} {The impact of school infrastructure on learning: A synthesis
  of the evidence.}
\newblock
\APACaddressPublisher{}{World Bank Publications}.
\newblock
\begin{APACrefURL} {\url{http://hdl.handle.net/10986/30920}} \end{APACrefURL}
\PrintBackRefs{\CurrentBib}

\bibitem [\protect \citeauthoryear {%
Besag%
}{%
Besag%
}{%
{\protect \APACyear {1974}}%
}]{%
CAR}
\APACinsertmetastar {%
CAR}%
\begin{APACrefauthors}%
Besag, J.%
\end{APACrefauthors}%
\unskip\
\newblock
\APACrefYearMonthDay{1974}{}{}.
\newblock
{\BBOQ}\APACrefatitle {Spatial interaction and the statistical analysis of
  lattice systems} {Spatial interaction and the statistical analysis of lattice
  systems}.{\BBCQ}
\newblock
\APACjournalVolNumPages{Journal of the Royal Statistical Society: Series B
  (Methodological)}{36}{2}{192--225,}
\newblock
\begin{APACrefDOI} \doi{10.1111/j.2517-6161.1974.tb00999.x} \end{APACrefDOI}
\newblock

\newblock

\PrintBackRefs{\CurrentBib}

\bibitem [\protect \citeauthoryear {%
Besag%
\ \BBA {} Kooperberg%
}{%
Besag%
\ \BBA {} Kooperberg%
}{%
{\protect \APACyear {1995}}%
}]{%
ICAR}
\APACinsertmetastar {%
ICAR}%
\begin{APACrefauthors}%
Besag, J.%
\BCBT {}\ \BBA {} Kooperberg, C.%
\end{APACrefauthors}%
\unskip\
\newblock
\APACrefYearMonthDay{1995}{12}{}.
\newblock
{\BBOQ}\APACrefatitle {On conditional and intrinsic autoregressions} {On
  conditional and intrinsic autoregressions}.{\BBCQ}
\newblock
\APACjournalVolNumPages{Biometrika}{82}{4}{733-746,}
\newblock
\begin{APACrefDOI} \doi{https://doi.org/10.1093/biomet/82.4.733}
  \end{APACrefDOI}
\newblock

\newblock

\PrintBackRefs{\CurrentBib}

\bibitem [\protect \citeauthoryear {%
Besag%
, York%
\BCBL {}\ \BBA {} Molli{\'e}%
}{%
Besag%
\ \protect \BOthers {.}}{%
{\protect \APACyear {1991}}%
}]{%
BYM}
\APACinsertmetastar {%
BYM}%
\begin{APACrefauthors}%
Besag, J.%
, York, J.%
\BCBL {} Molli{\'e}, A.%
\end{APACrefauthors}%
\unskip\
\newblock
\APACrefYearMonthDay{1991}{}{}.
\newblock
{\BBOQ}\APACrefatitle {Bayesian image restoration, with two applications in
  spatial statistics} {Bayesian image restoration, with two applications in
  spatial statistics}.{\BBCQ}
\newblock
\APACjournalVolNumPages{Annals of the institute of statistical
  mathematics}{43}{}{1--20,}
\newblock
\begin{APACrefDOI} \doi{https://doi.org/10.1007/BF00116466} \end{APACrefDOI}
\newblock

\newblock

\PrintBackRefs{\CurrentBib}

\bibitem [\protect \citeauthoryear {%
Bivand%
\ \protect \BOthers {.}}{%
Bivand%
\ \protect \BOthers {.}}{%
{\protect \APACyear {2017}}%
}]{%
spdep}
\APACinsertmetastar {%
spdep}%
\begin{APACrefauthors}%
Bivand, R.%
, Altman, M.%
, Anselin, L.%
, Assun{\c{c}}{\~a}o, R.%
, Berke, O.%
, Bernat, A.%
\BCBL {} Blanchet, G.%
\end{APACrefauthors}%
\unskip\
\newblock
\APACrefYearMonthDay{2017}{}{}.
\newblock
\APACrefbtitle {Package ‘spdep’: Spatial dependence: Weighting schemes,
  statistics, {R} package version.} {Package ‘spdep’: Spatial dependence:
  Weighting schemes, statistics, {R} package version.}
\newblock
\begin{APACrefURL} {\url{https://cran.r-project.org/package=spdep}}
  \end{APACrefURL}
\newblock
\APACrefnote{R package version 1.3-7}
\PrintBackRefs{\CurrentBib}

\bibitem [\protect \citeauthoryear {%
Blangiardo%
\ \BBA {} Cameletti%
}{%
Blangiardo%
\ \BBA {} Cameletti%
}{%
{\protect \APACyear {2015}}%
}]{%
Blangiardo}
\APACinsertmetastar {%
Blangiardo}%
\begin{APACrefauthors}%
Blangiardo, M.%
\BCBT {}\ \BBA {} Cameletti, M.%
\end{APACrefauthors}%
\unskip\
\newblock
\APACrefYear{2015}.
\newblock
\APACrefbtitle {Spatial and spatio-temporal {Bayesian} models with {R-INLA}}
  {Spatial and spatio-temporal {Bayesian} models with {R-INLA}}.
\newblock
\APACaddressPublisher{}{John Wiley \& Sons}.
\PrintBackRefs{\CurrentBib}

\bibitem [\protect \citeauthoryear {%
Bratti%
, Checchi%
\BCBL {}\ \BBA {} Filippin%
}{%
Bratti%
\ \protect \BOthers {.}}{%
{\protect \APACyear {2007}}%
}]{%
Bratti}
\APACinsertmetastar {%
Bratti}%
\begin{APACrefauthors}%
Bratti, M.%
, Checchi, D.%
\BCBL {} Filippin, A.%
\end{APACrefauthors}%
\unskip\
\newblock
\APACrefYearMonthDay{2007}{}{}.
\newblock
{\BBOQ}\APACrefatitle {Territorial differences in {Italian} students'
  mathematical competencies: evidence from Pisa 2003} {Territorial differences
  in {Italian} students' mathematical competencies: evidence from pisa
  2003}.{\BBCQ}
\newblock
\APACjournalVolNumPages{Giornale Degli Economisti e Annali Di
  Economia}{66}{3}{299--333,}
\newblock
\begin{APACrefURL} {\url{https://www.jstor.org/stable/23248253}}
  \end{APACrefURL}
\newblock

\newblock

\PrintBackRefs{\CurrentBib}

\bibitem [\protect \citeauthoryear {%
Bucci%
\ \protect \BOthers {.}}{%
Bucci%
\ \protect \BOthers {.}}{%
{\protect \APACyear {2023}}%
}]{%
BDI}
\APACinsertmetastar {%
BDI}%
\begin{APACrefauthors}%
Bucci, M.%
, Gazzano, L.%
, Gennari, E.%
, Grompone, A.%
, Ivaldi, G.%
, Messina, G.%
\BCBL {} Ziglio, G.%
\end{APACrefauthors}%
\unskip\
\newblock
\APACrefYearMonthDay{2023}{}{}.
\newblock
{\BBOQ}\APACrefatitle {Per chi suona la campan (ell) a? La dotazione di
  infrastrutture scolastiche in {Italia} (For whom the bell tolls? The
  availability of school infrastructure in {Italy})} {Per chi suona la campan
  (ell) a? la dotazione di infrastrutture scolastiche in {Italia} (for whom the
  bell tolls? the availability of school infrastructure in {Italy})}.{\BBCQ}
\newblock
\APACjournalVolNumPages{Politica economica}{}{}{1--50,}
\newblock
\begin{APACrefDOI} \doi{10.32057/0.QEF.2023.0827} \end{APACrefDOI}
\newblock

\newblock

\PrintBackRefs{\CurrentBib}

\bibitem [\protect \citeauthoryear {%
Cefalo%
, Pollice%
\BCBL {}\ \BBA {} Maranzano%
}{%
Cefalo%
\ \protect \BOthers {.}}{%
{\protect \APACyear {2024}}%
}]{%
SchoolDataIT}
\APACinsertmetastar {%
SchoolDataIT}%
\begin{APACrefauthors}%
Cefalo, L.%
, Pollice, A.%
\BCBL {} Maranzano, P.%
\end{APACrefauthors}%
\unskip\
\newblock
\APACrefYearMonthDay{2024}{}{}.
\newblock
\APACrefbtitle {{SchoolDataIT}: Retrieve, Harmonise and Map Open Data regarding
  the {Italian} School System.} {{SchoolDataIT}: Retrieve, harmonise and map
  open data regarding the {Italian} school system.}
\newblock
\begin{APACrefURL} {https://CRAN.R-project.org/package=SchoolDataIT}
  \end{APACrefURL}
\newblock
\APACrefnote{R Package Version 0.2.2}
\PrintBackRefs{\CurrentBib}

\bibitem [\protect \citeauthoryear {%
Cliff%
\ \BBA {} Ord%
}{%
Cliff%
\ \BBA {} Ord%
}{%
{\protect \APACyear {1981}}%
}]{%
Cliff_Ord}
\APACinsertmetastar {%
Cliff_Ord}%
\begin{APACrefauthors}%
Cliff, A.D.%
\BCBT {}\ \BBA {} Ord, J.K.%
\end{APACrefauthors}%
\unskip\
\newblock
\APACrefYear{1981}.
\newblock
\APACrefbtitle {Spatial Processes: Models and Applications} {Spatial processes:
  Models and applications}.
\newblock
\APACaddressPublisher{}{Pion, London}.
\PrintBackRefs{\CurrentBib}

\bibitem [\protect \citeauthoryear {%
Dupont%
, Marques%
\BCBL {}\ \BBA {} Kneib%
}{%
Dupont%
\ \protect \BOthers {.}}{%
{\protect \APACyear {2023}}%
}]{%
DupontArXiv}
\APACinsertmetastar {%
DupontArXiv}%
\begin{APACrefauthors}%
Dupont, E.%
, Marques, I.%
\BCBL {} Kneib, T.%
\end{APACrefauthors}%
\unskip\
\newblock
\APACrefYearMonthDay{2023}{}{}.
\newblock
\APACrefbtitle {Demystifying spatial confounding.} {Demystifying spatial
  confounding.}
\PrintBackRefs{\CurrentBib}

\bibitem [\protect \citeauthoryear {%
Dupont%
, Wood%
\BCBL {}\ \BBA {} Augustin%
}{%
Dupont%
\ \protect \BOthers {.}}{%
{\protect \APACyear {2022}}%
}]{%
Dupont}
\APACinsertmetastar {%
Dupont}%
\begin{APACrefauthors}%
Dupont, E.%
, Wood, S.N.%
\BCBL {} Augustin, N.H.%
\end{APACrefauthors}%
\unskip\
\newblock
\APACrefYearMonthDay{2022}{}{}.
\newblock
{\BBOQ}\APACrefatitle {Spatial+: a novel approach to spatial confounding}
  {Spatial+: a novel approach to spatial confounding}.{\BBCQ}
\newblock
\APACjournalVolNumPages{Biometrics}{78}{4}{1279--1290,}
\newblock
\begin{APACrefDOI} \doi{https://doi.org/10.1111/biom.13656} \end{APACrefDOI}
\newblock

\newblock

\PrintBackRefs{\CurrentBib}

\bibitem [\protect \citeauthoryear {%
Freni-Sterrantino%
, Ventrucci%
\BCBL {}\ \BBA {} Rue%
}{%
Freni-Sterrantino%
\ \protect \BOthers {.}}{%
{\protect \APACyear {2018}}%
}]{%
Freni}
\APACinsertmetastar {%
Freni}%
\begin{APACrefauthors}%
Freni-Sterrantino, A.%
, Ventrucci, M.%
\BCBL {} Rue, H.%
\end{APACrefauthors}%
\unskip\
\newblock
\APACrefYearMonthDay{2018}{}{}.
\newblock
{\BBOQ}\APACrefatitle {A note on intrinsic conditional autoregressive models
  for disconnected graphs} {A note on intrinsic conditional autoregressive
  models for disconnected graphs}.{\BBCQ}
\newblock
\APACjournalVolNumPages{Spatial and spatio-temporal
  epidemiology}{26}{}{25--34,}
\newblock
\begin{APACrefDOI} \doi{10.1016/j.sste.2018.04.002} \end{APACrefDOI}
\newblock

\newblock

\PrintBackRefs{\CurrentBib}

\bibitem [\protect \citeauthoryear {%
Geisser%
\ \BBA {} Eddy%
}{%
Geisser%
\ \BBA {} Eddy%
}{%
{\protect \APACyear {1979}}%
}]{%
Geisser}
\APACinsertmetastar {%
Geisser}%
\begin{APACrefauthors}%
Geisser, S.%
\BCBT {}\ \BBA {} Eddy, W.F.%
\end{APACrefauthors}%
\unskip\
\newblock
\APACrefYearMonthDay{1979}{}{}.
\newblock
{\BBOQ}\APACrefatitle {A predictive approach to model selection} {A predictive
  approach to model selection}.{\BBCQ}
\newblock
\APACjournalVolNumPages{Journal of the American Statistical
  Association}{74}{365}{153--160,}
\newblock
\begin{APACrefDOI} \doi{https://doi.org/10.1080/01621459.1979.10481632}
  \end{APACrefDOI}
\newblock

\newblock

\PrintBackRefs{\CurrentBib}

\bibitem [\protect \citeauthoryear {%
Gelfand%
, Dey%
\BCBL {}\ \BBA {} Chang%
}{%
Gelfand%
\ \protect \BOthers {.}}{%
{\protect \APACyear {1992}}%
}]{%
Gelfand}
\APACinsertmetastar {%
Gelfand}%
\begin{APACrefauthors}%
Gelfand, A.E.%
, Dey, D.K.%
\BCBL {} Chang, H.%
\end{APACrefauthors}%
\unskip\
\newblock
\APACrefYearMonthDay{1992}{}{}.
\newblock
{\BBOQ}\APACrefatitle {Model determination using predictive distributions with
  implementation via sampling-based methods} {Model determination using
  predictive distributions with implementation via sampling-based
  methods}.{\BBCQ}
\newblock
\APACjournalVolNumPages{Bayesian statistics 4}{}{}{147--168,}
\newblock
\begin{APACrefDOI} \doi{10.1093/oso/9780198522669.003.0009} \end{APACrefDOI}
\newblock

\newblock

\PrintBackRefs{\CurrentBib}

\bibitem [\protect \citeauthoryear {%
Gelfand%
\ \BBA {} Vounatsou%
}{%
Gelfand%
\ \BBA {} Vounatsou%
}{%
{\protect \APACyear {2003}}%
}]{%
PCAR_Gelfand}
\APACinsertmetastar {%
PCAR_Gelfand}%
\begin{APACrefauthors}%
Gelfand, A.E.%
\BCBT {}\ \BBA {} Vounatsou, P.%
\end{APACrefauthors}%
\unskip\
\newblock
\APACrefYearMonthDay{2003}{}{}.
\newblock
{\BBOQ}\APACrefatitle {Proper multivariate conditional autoregressive models
  for spatial data analysis} {Proper multivariate conditional autoregressive
  models for spatial data analysis}.{\BBCQ}
\newblock
\APACjournalVolNumPages{Biostatistics}{4}{1}{11--15,}
\newblock
\begin{APACrefDOI} \doi{https://doi.org/10.1093/biostatistics/4.1.11}
  \end{APACrefDOI}
\newblock

\newblock

\PrintBackRefs{\CurrentBib}

\bibitem [\protect \citeauthoryear {%
Gelman%
, Carlin%
, Stern%
\BCBL {}\ \BBA {} Rubin%
}{%
Gelman%
\ \protect \BOthers {.}}{%
{\protect \APACyear {2004}}%
}]{%
Gelman}
\APACinsertmetastar {%
Gelman}%
\begin{APACrefauthors}%
Gelman, A.%
, Carlin, J.B.%
, Stern, H.S.%
\BCBL {} Rubin, D.B.%
\end{APACrefauthors}%
\unskip\
\newblock
\APACrefYear{2004}.
\newblock
\APACrefbtitle {Bayesian Data Analysis} {Bayesian data analysis}\
  (\PrintOrdinal{2nd}\ \BEd).
\newblock
\APACaddressPublisher{}{Chapman and Hall/CRC}.
\PrintBackRefs{\CurrentBib}

\bibitem [\protect \citeauthoryear {%
Gelman%
, Hwang%
\BCBL {}\ \BBA {} Vehtari%
}{%
Gelman%
\ \protect \BOthers {.}}{%
{\protect \APACyear {2014}}%
}]{%
GelmanWAIC}
\APACinsertmetastar {%
GelmanWAIC}%
\begin{APACrefauthors}%
Gelman, A.%
, Hwang, J.%
\BCBL {} Vehtari, A.%
\end{APACrefauthors}%
\unskip\
\newblock
\APACrefYearMonthDay{2014}{}{}.
\newblock
{\BBOQ}\APACrefatitle {Understanding predictive information criteria for
  {Bayesian} models} {Understanding predictive information criteria for
  {Bayesian} models}.{\BBCQ}
\newblock
\APACjournalVolNumPages{Statistics and Computing}{24}{6}{997--1016,}
\newblock
\begin{APACrefDOI} \doi{10.1007/S11222-013-9416-2} \end{APACrefDOI}
\newblock

\newblock

\PrintBackRefs{\CurrentBib}

\bibitem [\protect \citeauthoryear {%
Giancola%
, Benadusi%
\BCBL {}\ \BBA {} Fornari%
}{%
Giancola%
\ \protect \BOthers {.}}{%
{\protect \APACyear {2010}}%
}]{%
Giancola}
\APACinsertmetastar {%
Giancola}%
\begin{APACrefauthors}%
Giancola, O.%
, Benadusi, L.%
\BCBL {} Fornari, R.%
\end{APACrefauthors}%
\unskip\
\newblock
\APACrefYearMonthDay{2010}{}{}.
\newblock
{\BBOQ}\APACrefatitle {Cos{\`\i} vicine, cos{\`\i} lontane. La questione
  dell’equit{\`a} scolastica nelle regioni italiane} {Cos{\`\i} vicine,
  cos{\`\i} lontane. la questione dell’equit{\`a} scolastica nelle regioni
  italiane}.{\BBCQ}
\newblock
\APACjournalVolNumPages{Scuola democratica}{1}{}{52--79,}
\newblock
\begin{APACrefURL} {\url{https://iris.uniroma1.it/handle/11573/339625}}
  \end{APACrefURL}
\newblock

\newblock

\PrintBackRefs{\CurrentBib}

\bibitem [\protect \citeauthoryear {%
Giancola%
\ \BBA {} Salmieri%
}{%
Giancola%
\ \BBA {} Salmieri%
}{%
{\protect \APACyear {2020}}%
}]{%
UniromaWP1}
\APACinsertmetastar {%
UniromaWP1}%
\begin{APACrefauthors}%
Giancola, O.%
\BCBT {}\ \BBA {} Salmieri, L.%
\end{APACrefauthors}%
\unskip\
\newblock
\APACrefYearMonthDay{2020}{}{}.
\newblock
\APACrefbtitle {Family Background, School-Track and Macro-Area: the Complex
  Chains of Education Inequalities in Italy.} {Family background, school-track
  and macro-area: the complex chains of education inequalities in italy.}
\newblock
\begin{APACrefURL} {\url{https://hdl.handle.net/11573/1366601}}
  \end{APACrefURL}
\newblock
\APACrefnote{working paper}
\PrintBackRefs{\CurrentBib}

\bibitem [\protect \citeauthoryear {%
Gómez-Rubio%
}{%
Gómez-Rubio%
}{%
{\protect \APACyear {2020}}%
}]{%
INLAbook}
\APACinsertmetastar {%
INLAbook}%
\begin{APACrefauthors}%
Gómez-Rubio, V.%
\end{APACrefauthors}%
\unskip\
\newblock
\APACrefYear{2020}.
\newblock
\APACrefbtitle {Bayesian Inference with {INLA}} {Bayesian inference with
  {INLA}}.
\newblock
\APACaddressPublisher{}{Chapman and Hall/CRC}.
\PrintBackRefs{\CurrentBib}

\bibitem [\protect \citeauthoryear {%
Held%
, Schrödle%
\BCBL {}\ \BBA {} Rue%
}{%
Held%
\ \protect \BOthers {.}}{%
{\protect \APACyear {2010}}%
}]{%
CPOINLA}
\APACinsertmetastar {%
CPOINLA}%
\begin{APACrefauthors}%
Held, L.%
, Schrödle, B.%
\BCBL {} Rue, H.%
\end{APACrefauthors}%
\unskip\
\newblock
\APACrefYearMonthDay{2010}{}{}.
\newblock
{\BBOQ}\APACrefatitle {Posterior and Cross-validatory Predictive Checks: A
  Comparison of {MCMC} and {INLA}} {Posterior and cross-validatory predictive
  checks: A comparison of {MCMC} and {INLA}}.{\BBCQ}
\newblock
 T.~Kneib\ \BBA {} G.~Tutz\ (\BEDS), \APACrefbtitle {Statistical Modelling and
  Regression Structures.} {Statistical modelling and regression structures.}
\newblock
\APACaddressPublisher{}{Physica-Verlag HD}.
\PrintBackRefs{\CurrentBib}

\bibitem [\protect \citeauthoryear {%
Hodges%
, Carlin%
\BCBL {}\ \BBA {} Fan%
}{%
Hodges%
\ \protect \BOthers {.}}{%
{\protect \APACyear {2003}}%
}]{%
Hodges2003}
\APACinsertmetastar {%
Hodges2003}%
\begin{APACrefauthors}%
Hodges, J.S.%
, Carlin, B.P.%
\BCBL {} Fan, Q.%
\end{APACrefauthors}%
\unskip\
\newblock
\APACrefYearMonthDay{2003}{}{}.
\newblock
{\BBOQ}\APACrefatitle {On the precision of the conditionally autoregressive
  prior in spatial models} {On the precision of the conditionally
  autoregressive prior in spatial models}.{\BBCQ}
\newblock
\APACjournalVolNumPages{Biometrics}{59}{2}{317--322,}
\newblock
\begin{APACrefDOI} \doi{10.1111/1541-0420.00038} \end{APACrefDOI}
\newblock

\newblock

\PrintBackRefs{\CurrentBib}

\bibitem [\protect \citeauthoryear {%
Hodges%
\ \BBA {} Reich%
}{%
Hodges%
\ \BBA {} Reich%
}{%
{\protect \APACyear {2010}}%
}]{%
Hodges}
\APACinsertmetastar {%
Hodges}%
\begin{APACrefauthors}%
Hodges, J.S.%
\BCBT {}\ \BBA {} Reich, B.J.%
\end{APACrefauthors}%
\unskip\
\newblock
\APACrefYearMonthDay{2010}{}{}.
\newblock
{\BBOQ}\APACrefatitle {Adding Spatially-Correlated Errors Can Mess Up the Fixed
  Effect You Love} {Adding spatially-correlated errors can mess up the fixed
  effect you love}.{\BBCQ}
\newblock
\APACjournalVolNumPages{The American Statistician}{64}{4}{325--334,}
\newblock
\begin{APACrefDOI} \doi{10.1198/tast.2010.10052} \end{APACrefDOI}
\newblock

\newblock

\PrintBackRefs{\CurrentBib}

\bibitem [\protect \citeauthoryear {%
{Infratel Italia}%
}{%
{Infratel Italia}%
}{%
{\protect \APACyear {2024}}%
}]{%
BB}
\APACinsertmetastar {%
BB}%
\begin{APACrefauthors}%
{Infratel Italia}%
\end{APACrefauthors}%
\unskip\
\newblock
\APACrefYearMonthDay{2024}{}{}.
\newblock
\APACrefbtitle {Schools Dashboard, part of the Ultra-Broadband Activation
  plan.} {Schools dashboard, part of the ultra-broadband activation plan.}
\newblock
\begin{APACrefURL}
  {\url{https://bandaultralarga.italia.it/scuole-voucher/progetto-scuole/}}
  \end{APACrefURL}
\newblock
\APACrefnote{last access December 19th 2024}
\PrintBackRefs{\CurrentBib}

\bibitem [\protect \citeauthoryear {%
{INVALSI - National Institute for the Evaluation of the Education System}%
}{%
{INVALSI - National Institute for the Evaluation of the Education System}%
}{%
{\protect \APACyear {2024}}%
}]{%
Invalsi_IS}
\APACinsertmetastar {%
Invalsi_IS}%
\begin{APACrefauthors}%
{INVALSI - National Institute for the Evaluation of the Education System}%
\end{APACrefauthors}%
\unskip\
\newblock
\APACrefYearMonthDay{2024}{}{}.
\newblock
\APACrefbtitle {Municipality Standardized Assessment Data (Invalsi Censuary
  Survey). Invalsi Statistical Services.} {Municipality standardized assessment
  data (invalsi censuary survey). invalsi statistical services.}
\newblock
\begin{APACrefURL}
  {\url{https://serviziostatistico.invalsi.it/invalsi_ss_data/dati-comunali-di-popolazione-comune-del-plesso/}}
  \end{APACrefURL}
\newblock
\APACrefnote{last access December 18th 2024}
\PrintBackRefs{\CurrentBib}

\bibitem [\protect \citeauthoryear {%
{ISTAT - Italian National Institute of Statistics}%
}{%
{ISTAT - Italian National Institute of Statistics}%
}{%
{\protect \APACyear {2022}}%
}]{%
InnerAreas}
\APACinsertmetastar {%
InnerAreas}%
\begin{APACrefauthors}%
{ISTAT - Italian National Institute of Statistics}%
\end{APACrefauthors}%
\unskip\
\newblock
\APACrefYearMonthDay{2022}{}{}.
\newblock
\APACrefbtitle {La geografia delle aree interne nel 2020: vasti territori tra
  potenzialità e debolezze.} {La geografia delle aree interne nel 2020: vasti
  territori tra potenzialità e debolezze.}
\newblock
\begin{APACrefURL}
  {\url{https://www.istat.it/it/files//2022/07/FOCUS-AREE-INTERNE-2021.pdf}}
  \end{APACrefURL}
\PrintBackRefs{\CurrentBib}

\bibitem [\protect \citeauthoryear {%
{Italian Ministry of Education, University and Research}%
}{%
{Italian Ministry of Education, University and Research}%
}{%
{\protect \APACyear {2024}}%
}]{%
MIUR}
\APACinsertmetastar {%
MIUR}%
\begin{APACrefauthors}%
{Italian Ministry of Education, University and Research}%
\end{APACrefauthors}%
\unskip\
\newblock
\APACrefYearMonthDay{2024}{}{}.
\newblock
\APACrefbtitle {Portale Unico dei Dati sulla scuola (Unique School Data
  Portal).} {Portale unico dei dati sulla scuola (unique school data portal).}
\newblock
\begin{APACrefURL} {\url{https://dati.istruzione.it/opendata/}}
  \end{APACrefURL}
\newblock
\APACrefnote{last accessed on December 17th 2024}
\PrintBackRefs{\CurrentBib}

\bibitem [\protect \citeauthoryear {%
{Italian Official Journal}%
}{%
{Italian Official Journal}%
}{%
{\protect \APACyear {2007}}%
}]{%
InvalsiLaw}
\APACinsertmetastar {%
InvalsiLaw}%
\begin{APACrefauthors}%
{Italian Official Journal}%
\end{APACrefauthors}%
\unskip\
\newblock
\APACrefYearMonthDay{2007}{}{}.
\newblock
\APACrefbtitle {Law 176 of October 25th 2007.} {Law 176 of october 25th 2007.}
\newblock
\begin{APACrefURL}
  {\url{https://www.gazzettaufficiale.it/atto/serie_generale/caricaDettaglioAtto/originario?atto.dataPubblicazioneGazzetta=2007-10-26&atto.codiceRedazionale=007G0195&elenco30giorni=false}}
  \end{APACrefURL}
\PrintBackRefs{\CurrentBib}

\bibitem [\protect \citeauthoryear {%
Khan%
\ \BBA {} Calder%
}{%
Khan%
\ \BBA {} Calder%
}{%
{\protect \APACyear {2022}}%
}]{%
Khan}
\APACinsertmetastar {%
Khan}%
\begin{APACrefauthors}%
Khan, K.%
\BCBT {}\ \BBA {} Calder, C.A.%
\end{APACrefauthors}%
\unskip\
\newblock
\APACrefYearMonthDay{2022}{}{}.
\newblock
{\BBOQ}\APACrefatitle {Restricted spatial regression methods: Implications for
  inference} {Restricted spatial regression methods: Implications for
  inference}.{\BBCQ}
\newblock
\APACjournalVolNumPages{Journal of the American Statistical
  Association}{117}{537}{482--494,}
\newblock
\begin{APACrefDOI} \doi{10.1080/01621459.2020.1788949} \end{APACrefDOI}
\newblock

\newblock

\PrintBackRefs{\CurrentBib}

\bibitem [\protect \citeauthoryear {%
Lamouroux%
, Geffroy%
, Leblond%
, Meyer%
\BCBL {}\ \BBA {} Albert%
}{%
Lamouroux%
\ \protect \BOthers {.}}{%
{\protect \APACyear {2024}}%
}]{%
Lamouroux}
\APACinsertmetastar {%
Lamouroux}%
\begin{APACrefauthors}%
Lamouroux, J.%
, Geffroy, A.%
, Leblond, S.%
, Meyer, C.%
\BCBL {} Albert, I.%
\end{APACrefauthors}%
\unskip\
\newblock
\APACrefYearMonthDay{2024}{}{}.
\newblock
{\BBOQ}\APACrefatitle {Addressing Spatial Confounding in geostatistical
  regression models: An {R-INLA} approach} {Addressing spatial confounding in
  geostatistical regression models: An {R-INLA} approach}.{\BBCQ}
\newblock
\APACjournalVolNumPages{arXiv e-prints}{}{}{arXiv:2410.01530,}
\newblock
\begin{APACrefDOI} \doi{10.48550/arXiv.2410.01530} \end{APACrefDOI}
\newblock
{\href{https://arxiv.org/abs/2410.01530}{{2410.01530}}}
\newblock

\PrintBackRefs{\CurrentBib}

\bibitem [\protect \citeauthoryear {%
Mardia%
}{%
Mardia%
}{%
{\protect \APACyear {1988}}%
}]{%
Mardia}
\APACinsertmetastar {%
Mardia}%
\begin{APACrefauthors}%
Mardia, K.%
\end{APACrefauthors}%
\unskip\
\newblock
\APACrefYearMonthDay{1988}{}{}.
\newblock
{\BBOQ}\APACrefatitle {Multi-dimensional multivariate {Gaussian} {Markov}
  random fields with application to image processing} {Multi-dimensional
  multivariate {Gaussian} {Markov} random fields with application to image
  processing}.{\BBCQ}
\newblock
\APACjournalVolNumPages{Journal of Multivariate Analysis}{24}{2}{265--284,}
\newblock
\begin{APACrefDOI} \doi{10.1016/0047-259X(88)90040-1} \end{APACrefDOI}
\newblock

\newblock

\PrintBackRefs{\CurrentBib}

\bibitem [\protect \citeauthoryear {%
Martini%
}{%
Martini%
}{%
{\protect \APACyear {2020}}%
}]{%
Invalsi2020}
\APACinsertmetastar {%
Invalsi2020}%
\begin{APACrefauthors}%
Martini, A.%
\end{APACrefauthors}%
\unskip\
\newblock
\APACrefYearMonthDay{2020}{}{}.
\newblock
\APACrefbtitle {Il Divario Nord-Sud nei Risultati delle Prove INVALSI.} {Il
  divario nord-sud nei risultati delle prove invalsi.}
\newblock
\APACrefnote{Invalsi Working Paper n. 52}
\PrintBackRefs{\CurrentBib}

\bibitem [\protect \citeauthoryear {%
Matteucci%
\ \BBA {} Mignani%
}{%
Matteucci%
\ \BBA {} Mignani%
}{%
{\protect \APACyear {2014}}%
}]{%
Matteucci}
\APACinsertmetastar {%
Matteucci}%
\begin{APACrefauthors}%
Matteucci, M.%
\BCBT {}\ \BBA {} Mignani, S.%
\end{APACrefauthors}%
\unskip\
\newblock
\APACrefYearMonthDay{2014}{}{}.
\newblock
{\BBOQ}\APACrefatitle {Exploring regional differences in the reading
  competencies of {Italian} students} {Exploring regional differences in the
  reading competencies of {Italian} students}.{\BBCQ}
\newblock
\APACjournalVolNumPages{Evaluation review}{38}{3}{251--290,}
\newblock
\begin{APACrefDOI} \doi{10.1177/0193841X14540289} \end{APACrefDOI}
\newblock

\newblock

\PrintBackRefs{\CurrentBib}

\bibitem [\protect \citeauthoryear {%
Nobre%
, Schmidt%
\BCBL {}\ \BBA {} Pereira%
}{%
Nobre%
\ \protect \BOthers {.}}{%
{\protect \APACyear {2021}}%
}]{%
Nobre}
\APACinsertmetastar {%
Nobre}%
\begin{APACrefauthors}%
Nobre, W.S.%
, Schmidt, A.M.%
\BCBL {} Pereira, J.B.%
\end{APACrefauthors}%
\unskip\
\newblock
\APACrefYearMonthDay{2021}{}{}.
\newblock
{\BBOQ}\APACrefatitle {On the effects of spatial confounding in hierarchical
  models} {On the effects of spatial confounding in hierarchical
  models}.{\BBCQ}
\newblock
\APACjournalVolNumPages{International Statistical Review}{89}{2}{302--322,}
\newblock
\begin{APACrefDOI} \doi{10.1111/insr.12407} \end{APACrefDOI}
\newblock

\newblock

\PrintBackRefs{\CurrentBib}

\bibitem [\protect \citeauthoryear {%
OECD%
}{%
OECD%
}{%
{\protect \APACyear {2009}}%
}]{%
PISA}
\APACinsertmetastar {%
PISA}%
\begin{APACrefauthors}%
OECD%
\end{APACrefauthors}%
\unskip\
\newblock
\APACrefYear{2009}.
\newblock
\APACrefbtitle {PISA Data Analysis Manual: SPSS, Second Edition} {Pisa data
  analysis manual: Spss, second edition}.
\newblock
\APACaddressPublisher{Paris}{OECD Publishing}.
\PrintBackRefs{\CurrentBib}

\bibitem [\protect \citeauthoryear {%
Palmí-Perales%
, Gómez-Rubio%
, Bivand%
, Cameletti%
\BCBL {}\ \BBA {} Rue%
}{%
Palmí-Perales%
\ \protect \BOthers {.}}{%
{\protect \APACyear {2023}}%
}]{%
INLAMSM2}
\APACinsertmetastar {%
INLAMSM2}%
\begin{APACrefauthors}%
Palmí-Perales, F.%
, Gómez-Rubio, V.%
, Bivand, R.S.%
, Cameletti, M.%
\BCBL {} Rue, H.%
\end{APACrefauthors}%
\unskip\
\newblock
\APACrefYearMonthDay{2023}{}{}.
\newblock
\APACrefbtitle {Bayesian Inference for Multivariate Spatial Models with
  {INLA}.} {Bayesian inference for multivariate spatial models with {INLA}.}
\newblock
\APACaddressPublisher{}{The R Journal}.
\PrintBackRefs{\CurrentBib}

\bibitem [\protect \citeauthoryear {%
Palmí-Perales%
, Gómez-Rubio%
\BCBL {}\ \BBA {} Martinez-Beneito%
}{%
Palmí-Perales%
\ \protect \BOthers {.}}{%
{\protect \APACyear {2021}}%
}]{%
INLAMSM}
\APACinsertmetastar {%
INLAMSM}%
\begin{APACrefauthors}%
Palmí-Perales, F.%
, Gómez-Rubio, V.%
\BCBL {} Martinez-Beneito, M.A.%
\end{APACrefauthors}%
\unskip\
\newblock
\APACrefYearMonthDay{2021}{}{}.
\newblock
{\BBOQ}\APACrefatitle {Bayesian Multivariate Spatial Models for Lattice Data
  with {INLA}} {Bayesian multivariate spatial models for lattice data with
  {INLA}}.{\BBCQ}
\newblock
\APACjournalVolNumPages{Journal of Statistical Software}{98}{2}{1--29,}
\newblock
\begin{APACrefDOI} \doi{10.18637/jss.v098.i02} \end{APACrefDOI}
\newblock

\newblock

\PrintBackRefs{\CurrentBib}

\bibitem [\protect \citeauthoryear {%
Pettit%
}{%
Pettit%
}{%
{\protect \APACyear {1990}}%
}]{%
CPO}
\APACinsertmetastar {%
CPO}%
\begin{APACrefauthors}%
Pettit, L.I.%
\end{APACrefauthors}%
\unskip\
\newblock
\APACrefYearMonthDay{1990}{}{}.
\newblock
{\BBOQ}\APACrefatitle {The Conditional Predictive Ordinate for the Normal
  Distribution} {The conditional predictive ordinate for the normal
  distribution}.{\BBCQ}
\newblock
\APACjournalVolNumPages{Journal of the Royal Statistical Society. Series B
  (Methodological)}{52}{1}{175--184,}
\newblock
\begin{APACrefDOI} \doi{10.1111/j.2517-6161.1990.tb01780.x} \end{APACrefDOI}
\newblock

\newblock

\PrintBackRefs{\CurrentBib}

\bibitem [\protect \citeauthoryear {%
Reich%
, Hodges%
\BCBL {}\ \BBA {} Zadnik%
}{%
Reich%
\ \protect \BOthers {.}}{%
{\protect \APACyear {2006}}%
}]{%
RHZ}
\APACinsertmetastar {%
RHZ}%
\begin{APACrefauthors}%
Reich, B.J.%
, Hodges, J.S.%
\BCBL {} Zadnik, V.%
\end{APACrefauthors}%
\unskip\
\newblock
\APACrefYearMonthDay{2006}{}{}.
\newblock
{\BBOQ}\APACrefatitle {Effects of residual smoothing on the posterior of the
  fixed effects in disease-mapping models} {Effects of residual smoothing on
  the posterior of the fixed effects in disease-mapping models}.{\BBCQ}
\newblock
\APACjournalVolNumPages{Biometrics}{62}{4}{1197--1206,}
\newblock
\begin{APACrefDOI} \doi{10.1111/j.1541-0420.2006.00617.x} \end{APACrefDOI}
\newblock

\newblock

\PrintBackRefs{\CurrentBib}

\bibitem [\protect \citeauthoryear {%
Rue%
, Martino%
\BCBL {}\ \BBA {} Chopin%
}{%
Rue%
\ \protect \BOthers {.}}{%
{\protect \APACyear {2009}}%
}]{%
INLA}
\APACinsertmetastar {%
INLA}%
\begin{APACrefauthors}%
Rue, H.%
, Martino, S.%
\BCBL {} Chopin, N.%
\end{APACrefauthors}%
\unskip\
\newblock
\APACrefYearMonthDay{2009}{}{}.
\newblock
{\BBOQ}\APACrefatitle {Approximate {Bayesian} inference for latent {Gaussian}
  models by using integrated nested {Laplace} approximations} {Approximate
  {Bayesian} inference for latent {Gaussian} models by using integrated nested
  {Laplace} approximations}.{\BBCQ}
\newblock
\APACjournalVolNumPages{Journal of the Royal Statistical Society Series B:
  (Methodological)}{71}{2}{319--392,}
\newblock
\begin{APACrefDOI} \doi{10.1111/j.1467-9868.2008.00700.x} \end{APACrefDOI}
\newblock

\newblock

\PrintBackRefs{\CurrentBib}

\bibitem [\protect \citeauthoryear {%
Silverman%
}{%
Silverman%
}{%
{\protect \APACyear {1986}}%
}]{%
Silverman}
\APACinsertmetastar {%
Silverman}%
\begin{APACrefauthors}%
Silverman, B.W.%
\end{APACrefauthors}%
\unskip\
\newblock
\APACrefYear{1986}.
\newblock
\APACrefbtitle {Density Estimation for Statistics and Data Analysis} {Density
  estimation for statistics and data analysis}.
\newblock
\APACaddressPublisher{London}{Chapman and Hall}.
\PrintBackRefs{\CurrentBib}

\bibitem [\protect \citeauthoryear {%
Simpson%
, Rue%
, Riebler%
, Martins%
\BCBL {}\ \BBA {} Sørbye%
}{%
Simpson%
\ \protect \BOthers {.}}{%
{\protect \APACyear {2017}}%
}]{%
PCprior}
\APACinsertmetastar {%
PCprior}%
\begin{APACrefauthors}%
Simpson, D.%
, Rue, H.%
, Riebler, A.%
, Martins, T.G.%
\BCBL {} Sørbye, S.H.%
\end{APACrefauthors}%
\unskip\
\newblock
\APACrefYearMonthDay{2017}{}{}.
\newblock
{\BBOQ}\APACrefatitle {Penalising Model Component Complexity: A Principled,
  Practical Approach to Constructing Priors} {Penalising model component
  complexity: A principled, practical approach to constructing priors}.{\BBCQ}
\newblock
\APACjournalVolNumPages{Statistical Science}{32}{1}{1 -- 28,}
\newblock
\begin{APACrefDOI} \doi{10.1214/16-STS576} \end{APACrefDOI}
\newblock

\newblock

\PrintBackRefs{\CurrentBib}

\bibitem [\protect \citeauthoryear {%
Spiegelhalter%
, Best%
, Carlin%
\BCBL {}\ \BBA {} Van Der~Linde%
}{%
Spiegelhalter%
\ \protect \BOthers {.}}{%
{\protect \APACyear {2002}}%
}]{%
DIC}
\APACinsertmetastar {%
DIC}%
\begin{APACrefauthors}%
Spiegelhalter, D.J.%
, Best, N.G.%
, Carlin, B.P.%
\BCBL {} Van Der~Linde, A.%
\end{APACrefauthors}%
\unskip\
\newblock
\APACrefYearMonthDay{2002}{10}{}.
\newblock
{\BBOQ}\APACrefatitle {Bayesian measures of model complexity and fit} {Bayesian
  measures of model complexity and fit}.{\BBCQ}
\newblock
\APACjournalVolNumPages{Journal of the Royal Statistical Society: Series B
  (Statistical Methodology)}{64}{4}{583-639,}
\newblock
\begin{APACrefDOI} \doi{10.1111/1467-9868.00353} \end{APACrefDOI}
\newblock

\newblock

\PrintBackRefs{\CurrentBib}

\bibitem [\protect \citeauthoryear {%
Sørbye%
\ \BBA {} Rue%
}{%
Sørbye%
\ \BBA {} Rue%
}{%
{\protect \APACyear {2014}}%
}]{%
Sorbye}
\APACinsertmetastar {%
Sorbye}%
\begin{APACrefauthors}%
Sørbye, S.%
\BCBT {}\ \BBA {} Rue, H.%
\end{APACrefauthors}%
\unskip\
\newblock
\APACrefYearMonthDay{2014}{}{}.
\newblock
{\BBOQ}\APACrefatitle {Scaling intrinsic {Gaussian} {Markov} random field
  priors in spatial modelling} {Scaling intrinsic {Gaussian} {Markov} random
  field priors in spatial modelling}.{\BBCQ}
\newblock
\APACjournalVolNumPages{Spatial Statistics}{8}{}{39--51,}
\newblock
\begin{APACrefDOI} \doi{10.1016/j.spasta.2013.06.004} \end{APACrefDOI}
\newblock

\newblock

\PrintBackRefs{\CurrentBib}

\bibitem [\protect \citeauthoryear {%
Urdangarin%
, Goicoa%
, Kneib%
\BCBL {}\ \BBA {} Ugarte%
}{%
Urdangarin%
\ \protect \BOthers {.}}{%
{\protect \APACyear {2024}}%
}]{%
Urdangarin24}
\APACinsertmetastar {%
Urdangarin24}%
\begin{APACrefauthors}%
Urdangarin, A.%
, Goicoa, T.%
, Kneib, T.%
\BCBL {} Ugarte, M.D.%
\end{APACrefauthors}%
\unskip\
\newblock
\APACrefYearMonthDay{2024}{}{}.
\newblock
{\BBOQ}\APACrefatitle {A simplified spatial+ approach to mitigate spatial
  confounding in multivariate spatial areal models} {A simplified spatial+
  approach to mitigate spatial confounding in multivariate spatial areal
  models}.{\BBCQ}
\newblock
\APACjournalVolNumPages{Spatial Statistics}{59}{}{100804,}
\newblock
\begin{APACrefDOI} \doi{10.1016/j.spasta.2023.100804} \end{APACrefDOI}
\newblock

\newblock

\PrintBackRefs{\CurrentBib}

\bibitem [\protect \citeauthoryear {%
Urdangarin%
, Goicoa%
\BCBL {}\ \BBA {} Ugarte%
}{%
Urdangarin%
\ \protect \BOthers {.}}{%
{\protect \APACyear {2023}}%
}]{%
Urdangarin23}
\APACinsertmetastar {%
Urdangarin23}%
\begin{APACrefauthors}%
Urdangarin, A.%
, Goicoa, T.%
\BCBL {} Ugarte, M.D.%
\end{APACrefauthors}%
\unskip\
\newblock
\APACrefYearMonthDay{2023}{}{}.
\newblock
{\BBOQ}\APACrefatitle {Evaluating recent methods to overcome spatial
  confounding} {Evaluating recent methods to overcome spatial
  confounding}.{\BBCQ}
\newblock
\APACjournalVolNumPages{Revista Matemática Complutense}{36}{2}{333--360,}
\newblock
\begin{APACrefDOI} \doi{10.1007/s13163-022-00449-8} \end{APACrefDOI}
\newblock

\newblock

\PrintBackRefs{\CurrentBib}

\bibitem [\protect \citeauthoryear {%
Van~Niekerk%
, Krainski%
, Rustand%
\BCBL {}\ \BBA {} Rue%
}{%
Van~Niekerk%
\ \protect \BOthers {.}}{%
{\protect \APACyear {2023}}%
}]{%
VBINLA}
\APACinsertmetastar {%
VBINLA}%
\begin{APACrefauthors}%
Van~Niekerk, J.%
, Krainski, E.%
, Rustand, D.%
\BCBL {} Rue, H.%
\end{APACrefauthors}%
\unskip\
\newblock
\APACrefYearMonthDay{2023}{}{}.
\newblock
{\BBOQ}\APACrefatitle {A new avenue for {Bayesian} inference with {INLA}} {A
  new avenue for {Bayesian} inference with {INLA}}.{\BBCQ}
\newblock
\APACjournalVolNumPages{Computational Statistics and Data Analysis}{181}{}{,}
\newblock
\begin{APACrefDOI} \doi{10.1016/j.csda.2023.107692} \end{APACrefDOI}
\newblock

\newblock

\PrintBackRefs{\CurrentBib}

\bibitem [\protect \citeauthoryear {%
Van~Niekerk%
\ \BBA {} Rue%
}{%
Van~Niekerk%
\ \BBA {} Rue%
}{%
{\protect \APACyear {2021}}%
}]{%
SNprior}
\APACinsertmetastar {%
SNprior}%
\begin{APACrefauthors}%
Van~Niekerk, J.%
\BCBT {}\ \BBA {} Rue, H.%
\end{APACrefauthors}%
\unskip\
\newblock
\APACrefYearMonthDay{2021}{}{}.
\newblock
{\BBOQ}\APACrefatitle {Skewed Probit Regression - Identifiability, Contraction
  and Reformulation} {Skewed probit regression - identifiability, contraction
  and reformulation}.{\BBCQ}
\newblock
\APACjournalVolNumPages{Revstat - Statistical Journal}{19}{}{1--22,}
\newblock
\begin{APACrefDOI} \doi{10.57805/revstat.v19i1.328} \end{APACrefDOI}
\newblock

\newblock

\PrintBackRefs{\CurrentBib}

\bibitem [\protect \citeauthoryear {%
Van~Niekerk%
\ \BBA {} Rue%
}{%
Van~Niekerk%
\ \BBA {} Rue%
}{%
{\protect \APACyear {2024}}%
}]{%
VB}
\APACinsertmetastar {%
VB}%
\begin{APACrefauthors}%
Van~Niekerk, J.%
\BCBT {}\ \BBA {} Rue, H.%
\end{APACrefauthors}%
\unskip\
\newblock
\APACrefYearMonthDay{2024}{}{}.
\newblock
{\BBOQ}\APACrefatitle {Low-rank variational {Bayes} correction to the {Laplace}
  method} {Low-rank variational {Bayes} correction to the {Laplace}
  method}.{\BBCQ}
\newblock
\APACjournalVolNumPages{The Journal of Machine Learning
  Research}{25}{62}{1--25,}
\newblock
\begin{APACrefURL} {\url{http://jmlr.org/papers/v25/21-1405.html}}
  \end{APACrefURL}
\newblock

\newblock

\PrintBackRefs{\CurrentBib}

\bibitem [\protect \citeauthoryear {%
Wang%
, Yue%
\BCBL {}\ \BBA {} Faraway%
}{%
Wang%
\ \protect \BOthers {.}}{%
{\protect \APACyear {2018}}%
}]{%
Wang}
\APACinsertmetastar {%
Wang}%
\begin{APACrefauthors}%
Wang, X.%
, Yue, Y.R.%
\BCBL {} Faraway, J.J.%
\end{APACrefauthors}%
\unskip\
\newblock
\APACrefYear{2018}.
\newblock
\APACrefbtitle {Bayesian Regression Modeling with {INLA}} {Bayesian regression
  modeling with {INLA}}.
\newblock
\APACaddressPublisher{}{Chapman and Hall/CRC}.
\PrintBackRefs{\CurrentBib}

\bibitem [\protect \citeauthoryear {%
Watanabe%
}{%
Watanabe%
}{%
{\protect \APACyear {2013}}%
}]{%
WAIC}
\APACinsertmetastar {%
WAIC}%
\begin{APACrefauthors}%
Watanabe, S.%
\end{APACrefauthors}%
\unskip\
\newblock
\APACrefYearMonthDay{2013}{}{}.
\newblock
{\BBOQ}\APACrefatitle {A widely applicable {Bayesian} information criterion} {A
  widely applicable {Bayesian} information criterion}.{\BBCQ}
\newblock
\APACjournalVolNumPages{The Journal of Machine Learning
  Research}{14}{1}{867--897,}
\newblock
\begin{APACrefURL} {\url{http://jmlr.org/papers/v14/watanabe13a.html}}
  \end{APACrefURL}
\newblock

\newblock

\PrintBackRefs{\CurrentBib}

\bibitem [\protect \citeauthoryear {%
Wickham%
}{%
Wickham%
}{%
{\protect \APACyear {2011}}%
}]{%
ggplot}
\APACinsertmetastar {%
ggplot}%
\begin{APACrefauthors}%
Wickham, H.%
\end{APACrefauthors}%
\unskip\
\newblock
\APACrefYearMonthDay{2011}{}{}.
\newblock
{\BBOQ}\APACrefatitle {ggplot2} {ggplot2}.{\BBCQ}
\newblock
\APACjournalVolNumPages{Wiley interdisciplinary reviews: computational
  statistics}{3}{2}{180--185,}
\newblock
\begin{APACrefURL}
  {\url{https://cran.r-project.org/web/packages/ggplot2/index.html}}
  \end{APACrefURL}
\newblock
\APACrefnote{R package version 3.5.1}
\newblock

\newblock

\PrintBackRefs{\CurrentBib}

\end{thebibliography}

\begin{appendices}

\section{Eigenvector removal} \label{Appendix:A}

\begin{figure}[htbp]
    \centering
    \includegraphics[width=0.9\textwidth]{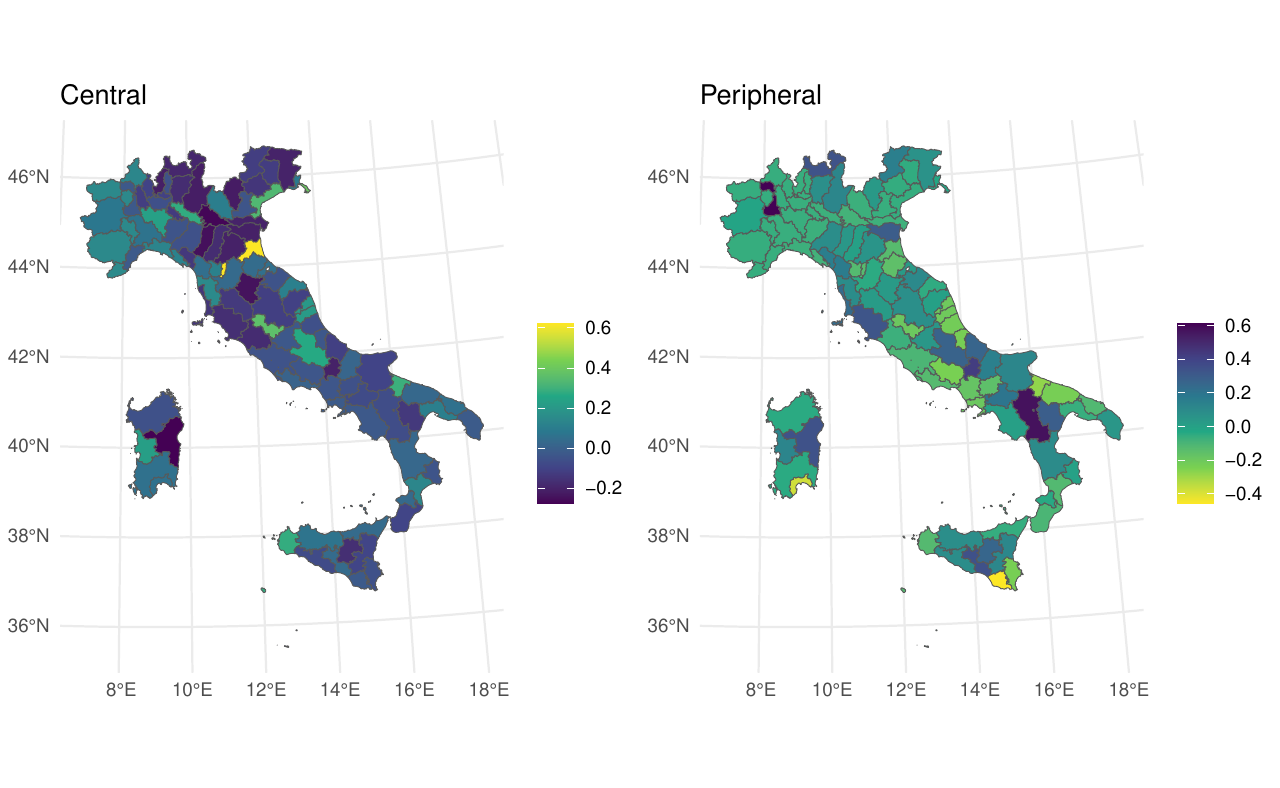} \\
    \caption{Proportion of central (left) and peripheral (right) municipalities per province once the spatial structure is removed by applying the S+(2) correction (see Table \ref{tab:eigenremoval}).}
    \label{fig:X_prov_nosp_1}
\end{figure}

In the following table eigenvector removal patterns are shown. The continent has 91 provinces and 186 infrastructural catchment areas; Sicily has 9 provinces and 13 infrastructural catchment areas; Sardinia has 5 provinces and 7 infrastructural catchment areas. The last eigenvector is constant within each component.
Patterns S+(1) and S+(3) are the simplest ones allowing to shrink the value of the standardised Moran's index below the $95$-th percentile of the Standard Normal distribution. At the province level, it is sufficient to remove the last 4 eigenvectors; at the level of infrastructural catchment areas this is only sufficient for the first two covariates, while more eigenvectors, corresponding to higher order trends, need to be removed for what concerns the other two variables. Pattern S+(2) allows, to the best of our findings, for the best ICAR fitting at the province level; S+(4) does the same at the level of infrastructural catchment areas; S+(5) and S+(6) serve the same purposes but for the PCAR model. In Figure \ref{fig:X_prov_nosp_1} the first two columns of $\mathbf{\bar{X}^{(NS)}}$ are shown, i.e. the province-level proportion of central and peripheral municipalities, using the S+(2) correction. Original values of these variables are in the upper panel of Figure \ref{fig:Xprov}.

\begin{table}[ht]
\begin{tabular}{lll|rrrr}
\toprule
Pattern & level & Component & Central & Peripheral & BB Activation & Urban transport \\
\midrule
\multirow{3}{*}{S+(1)} & \multirow{3}{*}{Prov} & Continent & 2 & 2 & 2 & 0 \\
                       &                       & Sicily & 1 & 1 & 1 & 0 \\
                       &                       & Sardinia & 1 & 1 & 1 & 0 \\
\multirow{3}{*}{S+(2)} & \multirow{3}{*}{Prov} & Continent & 5 & 4 & 5 & 0 \\
                       &                       & Sicily & 1 & 1 & 1 & 0 \\
                       &                       & Sardinia & 1 & 1 & 1 & 0 \\
\multirow{3}{*}{S+(3)} & \multirow{3}{*}{Pole} & Continent & 2 & 2 & 10$^{*}$ & 13 \\
                       &                       & Sicily & 1 & 1 & 1 & 1 \\
                       &                       & Sardinia & 1 & 1 & 1 & 1 \\
\multirow{3}{*}{S+(4)} & \multirow{3}{*}{Pole} & Continent & 8 & 8 & 9 & 10 \\
                       &                       & Sicily & 2 & 1 & 1 & 1 \\
                       &                       & Sardinia & 1 & 1 & 1 & 1 \\
\multirow{3}{*}{S+(5)} & \multirow{3}{*}{Prov} & Continent & 5 & 4 & 6 & 0 \\
                       &                       & Sicily & 1 & 1 & 1 & 0 \\
                       &                       & Sardinia & 1 & 1 & 1 & 0 \\
\multirow{3}{*}{S+(6)} & \multirow{3}{*}{Pole} & Continent & 9 & 8 & 7 & 13 \\
                       &                       & Sicily & 2 & 1 & 1 & 2 \\
                       &                       & Sardinia & 1 & 1 & 1 & 1 \\

\botrule
\end{tabular}
\caption{Eigenvectors removal patterns for each explanatory variable.$^{*}$ In this case, eigenvectors removed are the 172th and 178-186th ones}
\label{tab:eigenremoval}
\end{table}

\section{Extensive model comparison} \label{Appendix:B}
In Table \ref{tab:ICAR} all the models run throughout this analysis are compared. Two alternative strategies to approximate the full posterior of $\theta$ have been tested, namely the Variational Bayes and the Simplified Laplace approximations (respectively VB and SL, hereinafter). The former is implemented in the latest \texttt{R-INLA} framework and has been used to estimate the models whose results are summarised in Sections \ref{par:criteria} and \ref{section:results}. The latter allows to preserve information about skewness in the full conditional and is implemented in an older software framework. Due to the difficulties in locating the mode of $\Psi$, $y$ required to be centered at zero in the models approximated with the SL method.

Some additional model formulations are also compared: "NULL" denotes the model with no spatial effect (component-specific intercepts are still used); "IID" denotes the model with IID macroarea-level effects; "Ind. ICAR" denotes the model with two independent ICAR priors. The random intercept and the ICAR effect in the IID and independent ICAR models have an improper Uniform prior on the standard deviation. Models  S+(1), S+(2), S+(3), and S+(4)  are defined as in Appendix \ref{Appendix:A}. Alongside the selection criteria mentioned in Section \ref{par:criteria}, the two components of the DIC are shown, namely the expected deviance (Exp. Dev) and the effective number of parameters ($P_D$). The computational time of each model is shown as well, expressed in seconds. \\

This comparison highlights that the ICAR is more adequate to study Invalsi scores than both the null and IID models. In this latter case, point estimates are accurate (low MSE), but the high complexity suggests this may be due to overfitting. The correlated ICAR outperforms the independent one. The VB approximation also appears to be generally preferable over the SL.

\begin{table}[ht]
\centering
\begin{tabular}{rlllrrrrrrrr}
  \hline
  $z$ level & Approx & Model & -LPML & WAIC & DIC & Exp. Dev. & $P_D$ & MSE & time \\ 
  \toprule

  Null & VB & Null & 6979.502 & 13959.022 & 13959.264 & 13942.292 & \textbf{16.972} & 346.158 & 2.038 \\ 
  Prov & VB & IID & 6765.015 & 13524.273 & 13523.594 & 13368.734 & 154.860 & 233.906 & 6.204 \\ 
  Prov & VB & Ind. ICAR & 6722.306 & 13443.342 & 13443.291 & 13355.484 & 87.807 & 239.622 & 7.389 \\ 
  Prov & VB & Base ICAR & 6689.917 & 13379.149 & 13379.808 & 13307.056 & 72.752 & 235.551 & 11.501 \\ 
  Prov & VB & ICAR RSR & 6764.094 & 13526.761 & 13526.476 & 13440.811 & 85.665 & 251.886 & 10.197 \\ 
  Prov & VB & ICAR S+(1) & 6689.672 & 13378.651 & 13379.311 & 13306.071 & 73.239 & 235.326 & 11.111 \\ 
  Prov & VB & ICAR S+(2) & \textbf{6689.500} & \textbf{13378.303} & \textbf{13378.909} & 13305.406 & 73.502 & 235.189 & 11.613 \\ 
  Pole & VB & IID & 6840.664 & 13669.514 & 13666.691 & 13452.671 & 214.020 & 236.693 & 5.613 \\ 
  Pole & VB & Ind. ICAR & 6755.812 & 13508.925 & 13508.602 & 13390.671 & 117.931 & 240.193 & 17.936 \\ 
  Pole & VB & Base ICAR & 6694.435 & 13387.776 & 13388.726 & 13300.529 & 88.197 & 232.320 & 12.178 \\ 
  Pole & VB & ICAR RSR & 6754.037 & 13505.502 & 13507.315 & 13388.609 & 118.706 & 239.313 & 14.593 \\ 
  Pole & VB & ICAR S+(3) & 6694.443 & 13387.747 & 13388.711 & \textbf{13298.501} & 90.210 & 231.811 & 13.932 \\ 
  Pole & VB & ICAR S+(4) & 6694.055 & 13387.018 & 13387.948 & 13298.904 & 89.044 & 231.838 & 13.276 \\ 
  \midrule
  Null & SL & Null & 6979.612 & 13959.098 & 13959.450 & 13942.372 & 17.078 & 346.156 & 4.137 \\ 
  Prov & SL & IID & 6776.608 & 13524.444 & 13525.161 & 13369.837 & 155.324 & 234.058 & 11.304 \\ 
  Prov & SL & Ind. ICAR & 6725.692 & 13444.008 & 13444.536 & 13355.918 & 88.619 & 239.603 & 14.828 \\ 
  Prov & SL & Base ICAR & 6691.699 & 13379.533 & 13380.625 & 13307.040 & 73.585 & 235.470 & 24.948 \\ 
  Prov & SL & ICAR RSR & 6769.474 & 13527.510 & 13527.747 & 13441.562 & 86.184 & 251.937 & 40.036 \\ 
  Prov & SL & ICAR S+(1) & 6691.473 & 13379.438 & 13380.316 & 13307.059 & 73.257 & 235.465 & 24.181 \\ 
  Prov & SL & ICAR S+(2) & 6691.157 & 13379.047 & 13380.227 & 13306.657 & 73.570 & 235.362 & 33.475 \\ 
  Pole & SL & IID & 6857.701\footnotemark[1]  & 13669.789 & 13669.272 & 13453.829 & 215.442 & 236.662 & 13.597 \\ 
  Pole & SL & Ind. ICAR & 6764.696 & 13509.405 & 13509.590 & 13391.686 & 117.904 & 240.333 & 20.426 \\ 
  Pole & SL & Base ICAR & 6696.355 & 13388.114 & 13389.168 & 13300.653 & 88.515 & 232.351 & 31.058 \\ 
  Pole & SL & ICAR RSR & 6761.415 & 13506.393 & 13508.790 & 13389.828 & 118.962 & 239.459 & 54.796 \\ 
  Pole & SL & ICAR S+(3) & 6695.756 & 13388.190 & 13389.662 & 13298.914 & 90.748 & \textbf{231.808} & 31.176 \\  
  Pole & SL & ICAR S+(4) & 6696.531 & 13387.640 & 13388.553 & 13299.852 & 88.701 & 232.006 & 33.554 \\ 
    \botrule \end{tabular}
\caption{Model diagnostics for all the combinations of approximation approach to $\pi(\theta | y)$, aggregation level of $z$, and model employed for $z$, either null ($z$ not included, nonspatial model), IID (random intercept), independent bivariate ICAR, or dependent bivariate ICAR under either the base formulation, RSR or Spatial+2.0. Models are compared through negative Log Posterior Marginal Likelihood, Watanabe-Akaike information criterion, Deviance Information Criterion, expected deviance, effective number of parameters, Mean Square Error of posterior predictive response and computational time in seconds.}
\footnotetext[1] {Model re-run to avoid unreliable approximations to the CPOs.}
\label{tab:ICAR}
\end{table}

The value of Conditional Predictive Ordinates (CPOs) computed by \texttt{R-INLA} is reliable only under some regularity conditions \citep{CPOINLA}, which are always met except for the model with IID latent effects defined for infrastructural catchment areas and computed using the SL approximation; in this case, the observation for which the CPO computation is not reliable corresponds to the municipality of Melzo (MI), having the record highest Italian score (230 points). The CPOs for that model have then been recompiled (function \texttt{\detokenize{INLA::inla.cpo()}}). 
Though the results are quite similar (values of all selection criteria are slightly lower under the VB approximation), the SL approximation required to center $y$ at zero, other than being less computationally efficient. This encourages to employ the latest \texttt{R-INLA} version supporting the VB approximation even if we have a skewed likelihood for one of the two target variables.

\subsection{Estimates of $\beta$ under the nonspatial model and under Restricted Regression} \label{Appendix:null_RSR}

In Table \ref{tab:null_RSR} the estimated covariate effects under the model with no spatial effect (nonspatial) and under RSR is shown, with $z$ defined across provinces. Both models are estimated with the VB correction to the Gaussian approximation ("new" \texttt{R-INLA} version).
For what concerns the nonspatial model, explaining $y$ without a spatial latent field has the effect of raising estimates of $\beta$ with respect to the (unrestricted) spatial model.

\begin{table}[ht]
\centering
\begin{tabular}{ll|rrrr|rrrr}
  \hline
  && \multicolumn{4}{c|}{Nonspatial model} & \multicolumn{4}{c}{RSR model}\\
 & Subj & mean & sd & LB & UB & mean & sd & LB & UB \\ 
  \hline
  Central & MAT & 4.340 & 1.116 & 2.151 & 6.530 & 4.347 & 0.944 & 2.496 & 6.198 \\ 
  Central & ITA & 3.656 & 1.109 & 1.489 & 5.839 & 3.513 & 0.988 & 1.583 & 5.458 \\ 
  Peripheral & MAT & -5.055 & 1.205 & -7.418 & -2.692 & -5.035 & 1.019 & -7.034 & -3.036 \\ 
  Peripheral  & ITA & -4.092 & 1.161 & -6.368 & -1.813 & -4.090 & 1.028 & -6.105 & -2.072 \\ 
  BB Activation & MAT & 4.605 & 1.283 & 2.088 & 7.121 & 4.622 & 1.085 & 2.493 & 6.750 \\ 
  BB Activation & ITA & 3.564 & 1.225 & 1.168 & 5.974 & 3.346 & 1.102 & 1.190 & 5.512 \\ 
  Urban transport & MAT & 4.281 & 1.160 & 2.005 & 6.556 & 4.310 & 0.981 & 2.385 & 6.235 \\ 
  Urban transport & ITA & 3.744 & 1.126 & 1.539 & 5.957 & 4.017 & 0.991 & 2.075 & 5.964 \\ 
   \hline
\end{tabular}
\caption{Posterior summaries of covariate effects under the nonspatial model and the RSR-ICAR model defined at the province level}
\label{tab:null_RSR}
\end{table}

\section{Extension: the proper CAR model} \label{Appendix:C}
The bivariate proper CAR \citep{PCAR_Gelfand} extends the ICAR described in Section \ref{par:ICAR} by introducing an additional scalar parameter $\phi \in ]0, 1[$ which measures the strength of spatial association, the limit case of $\phi = 1$ being the ICAR. The distribution of a generic $z_i$ conditioned on all other sites is:
\begin{equation}
z_{i}   |  z_{-i}, \Lambda \sim N \left(\sum_{j \sim i} \phi \frac{w_{ij}}{d_i} z_{j},  \, \frac{1}{d_i} \Lambda^{-1}\right)
\label{eq:PCAR_local}
\end{equation}
which  implies a joint proper Normal distribution with mean $0$ and nonsingular precision matrix $\Lambda \otimes (\mathbf{D} - \phi \mathbf{W})$ with notation akin to Section \ref{par:ICAR}.
%
%
 Since no constraint is needed on the spatial effect, a unique intercept is employed. A Uniform prior is assigned to $\phi$, while all other ones are those described in Section \ref{section:Model_outline} with the exception of $\Lambda$, which is assigned a $\mathrm{Wishart}_k(I_k, k)$ prior \citep{INLAMSM}, with $k=2$. Differently from the ICAR model, here the precision matrix $\Lambda \otimes (\mathbf{D} - \phi \mathbf{W})$ depends on an unknown parameter, hence it is not possible to scale it a priori. Models comparison is show in Table \ref{tab:PCAR}, in analogy with Table \ref{tab:ICAR}. 

\begin{table}[ht]
\centering
\begin{tabular}{llrrrrrrr}
  \hline
  $z$ level & Model & -LPML & WAIC & DIC & Exp. Dev. & $P_D$ & MSE & time \\ 
  \hline  
  Prov & Base & 6688.512 & 13376.161 & 13376.216 & 13298.161 & 78.055 & 233.765 & 17.665 \\ 
  Prov & RSR &  6819.103 & 13636.395 & 13635.497 & 13544.723 & 90.774 & 265.933 & 23.433 \\ 
  Prov & S+(1) & 6688.291 & 13375.719 & 13375.782 & 13297.530 & 78.251 & 233.641 & 18.764 \\ 
  Prov & S+(5) & 6688.174 & 13375.485 & 13375.591 & 13297.392 & 78.199 & 233.613 & 18.381 \\  
  Pole & Base &  6694.183 & 13387.107 & 13387.811 & 13292.768 & 95.044 & 230.458 & 23.675 \\ 
  Pole & RSR & 6808.196 & 13613.277 & 13615.525 & 13486.800 & 128.724 & 250.825 & 23.227 \\ 
  Pole & S+(3)  & 6694.396 & 13387.495 & 13388.167 & 13291.686 & 96.481 & 230.131 & 22.565 \\ 
  Pole & S+(6) & 6693.845 & 13386.428 & 13387.251 & 13291.933 & 95.318 & 230.136 & 24.635 \\
   \hline
\end{tabular}
\caption{Model diagnostics for 8 proper CAR formulations: spatial aggregation level of $z$, spatial confounding treatment, negative Log Pseudo Marginal Likelihood, Watanabe-Akaike Information criterion, Deviance Information Criterion, expected deviance, effective number of parameters, Mean Squared Error of posterior predictive responses and computational time in seconds.}
\label{tab:PCAR}
\end{table}
S+(5) and S+(6) are the modified Spatial+ models which, to the best of our findings, have the lowest WAIC (see Table \ref{tab:eigenremoval}), with $z$ defined respectively at the province  and at the infrastructural catchment area level. Compared to Table \ref{tab:ICAR_diagnostics}, a general improvement in selection criteria can be noticed, which makes the proper CAR an appealing alternative for this kind of analysis.

In Tables \ref{tab:PCARfix} and \ref{tab:PCARhyper} posterior summaries are shown for covariate effects and hyperparameters respectively, both under the base and the S+(5) formulations. Variance parameters $\sigma_1^2$ and $\sigma_2^2$ (diagonal entries in $\Lambda^{-1}$) are however less easy to interpret since precision is not scaled.

Results are similar to the ICAR, which is consistent with the finding that $\phi$ ranges close to one, denoting that spatial association is strong. 

\begin{table}[ht]
\centering
\begin{tabular}{ll|rrrr|rrrr}
  \hline
  && \multicolumn{4}{c|}{Base model} & \multicolumn{4}{c}{S+(5)}\\
 & Subj & mean & sd & LB & UB & mean & sd & LB & UB \\ 
  \hline
  Intercept & MAT & 190.232 & 3.390 & 183.366 & 196.999 & 191.948 & 3.735 & 184.338 & 199.394 \\ 
  Intercept & ITA & 186.155 & 2.844 & 180.389 & 191.818 & 187.472 & 3.102 & 181.139 & 193.645 \\  
  Central & MAT& 2.721 & 0.909 & 0.938 & 4.504 & 2.507 & 0.890 & 0.762 & 4.252 \\
  Central & ITA  & 2.370 & 0.993 & 0.430 & 4.324 & 2.441 & 0.976 & 0.534 & 4.361 \\ 
  Peripheral & MAT & -2.299 & 1.005 & -4.270 & -0.327 & -2.057 & 0.958 & -3.936 & -0.177 \\ 
  Peripheral & ITA & -1.906 & 1.046 & -3.955 & 0.147 & -1.844 & 0.998 & -3.801 & 0.116 \\ 
  BB Activation& MAT & 3.320 & 1.076 & 1.209 & 5.430 & 3.285 & 1.052 & 1.222 & 5.348 \\ 
  BB Activation & ITA & 2.324 & 1.126 & 0.119 & 4.537 & 2.108 & 1.101 & -0.048 & 4.272 \\ 
  Urban transport & MAT & 2.529 & 1.054 & 0.462 & 4.595 & 2.566 & 1.053 & 0.500 & 4.632 \\ 
  Urban transport & ITA & 2.892 & 1.068 & 0.801 & 4.989 & 2.885 & 1.067 & 0.796 & 4.980 \\    \hline
\end{tabular}
\caption{Posterior summaries of intercepts and covariate effects when $z$ is defined as a province-level PCAR, under the base model and the Spatial+2.0 correction with removal pattern S+(5)}
\label{tab:PCARfix}
\end{table}

\begin{table}[ht]
\centering
\begin{tabular}{ll|rrrr|rrrr}
  \hline
  && \multicolumn{4}{c|}{Base model} & \multicolumn{4}{c}{Spatial+}\\
 & Subj & LB & median & UB & sd & LB & median & UB & sd \\ 
  \hline
  $\phi$ &         & 0.960 & 0.987 & 0.997 & 0.010 & 0.962 & 0.989 & 0.997 & 0.009 \\ 
  $\sigma_1^2$ & MAT & 30.028 & 48.739 & 79.074 & 12.552 & 30.429 & 49.150 & 79.739 & 12.619 \\ 
  $\sigma_2^2$ & ITA & 18.911 & 31.903 & 53.761 & 8.927 & 19.183 & 32.134 & 53.706 & 8.841 \\ 
  $\rho$ &      & 0.887 & 0.968 & 0.993 & 0.028 & 0.882 & 0.971 & 0.993 & 0.030 \\ 
  $\omega_1$ & MAT & 100.290 & 110.382 & 121.453 & 5.389 & 100.267 & 110.319 & 121.408 & 5.384 \\ 
  $\omega_2$ & ITA & 118.984 & 131.322 & 145.155 & 6.665 & 118.966 & 131.250 & 145.128 & 6.663 \\ 
  $\gamma_1$ &       & -0.499 & -0.375 & -0.237 & 0.067 & -0.496 & -0.373 & -0.235 & 0.066 \\ 
   \hline
\end{tabular}
\caption{Posterior summaries of hyperparameters with $z$ defined as a province-level PCAR, under the base model and the Spatial+2.0 correction with removal pattern S+(5)}
\label{tab:PCARhyper}
\end{table}

\end{appendices}

\end{document}